\documentclass{article}
\usepackage{arxiv}
\usepackage{framed,multirow}
\usepackage{moreverb}
\usepackage{graphicx}
\usepackage[vlined,boxed,ruled]{algorithm2e}
\usepackage{amsmath}
\usepackage{rotating}
\usepackage{multirow}
\usepackage{pict2e}
\usepackage{makecell}
\usepackage{slashbox}
\usepackage{indentfirst}
\usepackage[colorlinks,bookmarksopen,bookmarksnumbered,citecolor=red,urlcolor=red]{hyperref}
\usepackage{subfigure}
\usepackage{harpoon}
\usepackage{booktabs}
\usepackage[utf8]{inputenc} 
\usepackage[T1]{fontenc}    
\usepackage{hyperref}       
\usepackage{url}            
\usepackage{booktabs}       
\usepackage{amsfonts}       
\usepackage{nicefrac}       
\usepackage{microtype}      
\usepackage{lipsum}
\usepackage{graphicx}
\usepackage{float}
\usepackage{epstopdf}
\usepackage{amssymb}
\usepackage{latexsym}

\newtheorem{df}{Definition}
\usepackage{url}
\usepackage{xcolor}
\graphicspath{ {./images/} }

\title{Online Encrypted Skype Identification Based on an Updating Mechanism}

\author{
 Shi Dong \\
  School of Computer Science and Technology \\
  Zhoukou Normal University\\
  Zhoukou, China 466001 \\
  \texttt{dongshi@zknu.edu.cn} \\
}

\begin{document}
\maketitle
\begin{abstract}
The machine learning algorithm is gaining prominence in traffic identification research as it offers a way to overcome the shortcomings of port-based and deep packet inspection, especially for P2P-based Skype. However,recent studies have focused mainly on traffic identification based on a full-packet dataset, which poses great challenges to identifying online network traffic. This study aims to provide a new flow identification algorithm by taking the sampled flow records as the object. The study constructs flow records from a Skype set as the dataset, considers the inherent NETFLOW and extended flow metrics as features, and uses a fast correlation-based filter algorithm to select highly correlated features. The study also proposes a new NFI method that adopts a Bayesian updating mechanism to improve the classifier model. The experimental results show that the proposed scheme can achieve much better identification performance than existing state-of-the-art traffic identification methods, and a typical feature metric is analyzed in the sampling environment. The NFI method improves identification accuracy and reduces false positives and false negatives compared to other methods.
\end{abstract}




\section{Introduction}
\label{sec:introduction}
With growth in network bandwidth, network behavior patterns have become increasingly complex and led to an array of new network applications. Substantial attention is now focused on network traffic identification in the field of network management. Skype flow identification is an important foundation for network policy billing and differentiated services. As Skype uses a private communication protocol, and communication content is encrypted between users and the Skype server or among users, a detection method based on ports and features makes it difficult to effectively identify Skype traffic. Recent research on Skype flow identification focuses mostly on characteristics and communication mechanisms. M Korczynski \emph{et al}. \cite{Moore2005Toward} proposed stochastic fingerprints for application traffic flows conveyed in Secure Socket Layer/Transport Layer Security (SSL/TLS) sessions. S Molnr \emph{et al}.\cite{Kumar2006Advanced} suggested identification methods allowing us to discover logged on Skype users and their voice calls. Adami \emph{et al}.\cite{Nguyen2009A} put forth a real-time algorithm (named Skype-Hunter) to detect and classify Skype traffic. L Lu \emph{et al}.\cite{Dainotti2012Issues} studied inbound and outbound flow characteristics of the network end node and P2P feature to identify Skype traffic; however, the method must meet the requirement that the network topology endpoint is known, and single-ended node traffic can be obtained. Yet these conditions are challenging to fulfill in a realistic network environment, rendering the utility of the method limited. D Bonfiglio \emph{et al}.\cite{buda2018systematic} pointed out that Skype relied on two different modes to transfer VoIP application data: one is end-to-end(E2E), in which VoIP data are transferred between two end nodes; and end-to-phone (E2P), in which data are transferred between the end node and traditional PSTN telephone. The literature generally adopts the $\chi$ Square classifier to identify the encrypted network traffic and then uses a Bayesian classifier combined with real-time flow features to identify Skype flow. The algorithm does not consider that Skype P2P features and network traffic are smaller in the experimental network environment; therefore, the experimental results are difficult to explain fully with this model. Z Yuan \emph{et al}.\cite{Shi2013Research} first revealed the unique sequence signatures of Skype UDP flows and then implemented a practical online system called SkyTracer for precise Skype traffic identification. Author links open overlay panel. Naive Bayesian classification (NBC) has been widely applied in traffic identification as noted by \cite{Moore2005Toward}. This study aims to improve NBC to address the problem of identification accuracy.\\
The rest of this paper is structured as follows. Section 2 presents an overview of related work followed by machine learning algorithms for traffic classification. Section 3 proposes a traffic classification model based on extended NETFLOW flow metrics. Section 4 describes the proposed evaluation method and analyzes the experimental results. Finally, conclusions are drawn in Section 5.

\section{Related work}
The goal of machine learning is to identify sample data and build a learning classifier,then classify the testing samples through the constructed classifier. Machine learning has been applied to network traffic identification to solve the problem of deep packet inspection methods being unable to identify encrypted traffic. Common methods include NaiveBayes, BayesNet, and others. L jun \emph{et al}.\cite{Shi2013Flow} extracted the relevant feature characteristics and used genetic algorithms to select features and then adopted the Bayesian network to identify P2P traffic. Experiments showed that the K2, TAN, and BAN achieved better classification accuracy and faster classification speed than prior methods. However, this learning approach is probability-based and overly dependent on the distribution of the sample space. X peng \emph{et al}.\cite{Moore2005Internet} proposed a support vector machine (SVM) algorithm and compared it with an NB, NBK, NB+FCBF, and NBK+FCBF algorithm. Experimental results revealed that the overall accuracy of SVM without a feature selection algorithm was better than that of NB, slightly better than the NBK+FCBF algorithm using two kinds of optimization strategies, and effectively avoided the impact caused by unstable factors, providing an obvious advantage in dealing with the traffic classification problem. J zhang \emph{et al}.\cite{Zhu2007Accurate} presented a novel traffic classification scheme to improve classification performance when few training data were available. Experimental results indicated that the proposed scheme achieved much better classification performance than existing state-of-the-art traffic classification methods. L peng \emph{et al}. \cite{Williams2006A} first applied mutual information to analyze the information the first n packets provide to the flow type; their experimental dataset also included Skype. T qin \emph{et al}.\cite{Provost2008Machine} employed the signature of Packet Size Distribution (PSD) to capture flow dynamics, which is defined as the payload length distribution probability of the packets in one Bi-flow. S Mongkolluksamee \emph{et al}. \cite{Charte2015Addressing} suggested combining the packet size distribution and communication patterns extracted graphically to identify a mobile application. Validation results from five popular mobile applications (Facebook, Line, Skype, YouTube, and Web) demonstrated that the combined method realized high performance (0.95) of the F-measure even when using only 50 randomly sampled packets during a 3-minute interval. However, the timeliness problem of the dataset remains unresolved; it exerts a differential impacts on traffic identification and reduces identification accuracy. Mario et al. \cite{di2018improving} propose an enhanced detection method of encrypted Skype traffic by using an ad-hoc developed enhanced probe (ESkyPRO) which correlate the information received by ESkyPRO and other types of data obtained by an Intrusion Detection System (IDS) probe. Tian et al.\cite{tian2019complex} put forwards to the effective behavior model, which can effectively identify complex applications. Huang et al.\cite{huang2021complex} chooses Skype as the analysis object and propose the complex private network mining technology, which can solve many difficulties and challenges of CA identification. Qiao et al.\cite{qiao2021encrypted} propose to employ Long Short-Term Memory (LSTM) and Convolutional Neural Network (CNN) to classify encrypted 5G VPN voice traffic. However, above methods still can not effectively improve the identification accuracy. I also do some research work\cite{dong2021traffic,dong2021multi,dong2021network,dong2019traffic,dong2022network} for traffic identification. But research object is not encrypted skype traffic.
\section{Netflow flow identification}
This section presents a novel NB-based classification scheme, netflow flow identification (NFI), to effectively address the timeliness problem of the dataset and significantly improve classification performance even with a small set of supervised training data. First, the Bayes classifier will be introduced followed by the Bayes-updating model and algorithm.
\subsection{Bayes classifier}
According to the principle of machine learning methods and the Bayesian formula, this paper first outlines Bayesian network traffic classification.
\begin{df}
For network traffic application type $c=(c_1,c_2,...,c_i,...,c_n)$ and any network traffic flow x,$X=\{x_1,x_2,x_3,...,x_t\}$,according to Bayes theory, the following conditional probability \\

\end{df}
\begin{equation}
p(c_j|x)=\frac{p(x|c_j).p(c_j)}{\sum_{j=1}^n p(c_j).p(x|c_j)}
\end{equation}
where $p(c_j)$is the priori probability of $c_j$;$p(x|c_j)$ is the conditional probability of flow x when $c_j$ is known. Because the flow feature $A={A_1,A_2,...,A_k}$ can be expressed as network flow x; the Bayesian classifier assumes that the network flow feature vector A is independent and follows the Gaussian distribution. Therefore, the conditional probability of network flow x belonging to the class $c_j$  should meet the following condition:
\begin{equation}
p(x|c_j)=\prod_{i=1}^k p(A_i|c_j)
\end{equation}
such that Formula 1 can be deduced as follows:
\begin{equation}
p(c_j|A_1,A_2,...,A_k)=\frac{\prod_{i=1}^k p(A_i|c_j).p(c_j)}{\sum_{j=1}^n \prod_{i=1}^k p(A_i|c_j).p(c_j)}
\end{equation}

In the actual network traffic classification, the mutual independence between the characteristic feature of the network flow and the Gaussian distribution assumption is not entirely accurate; for example, the length of all network data equals the sum of the network length of the header portion and payload length of the network data. These characteristic attributes are mutually dependent but do not fully meet the independence assumption. Therefore, this study adopts the fast correlation-based filter (FCFB) feature selection algorithm to include as many network flow characteristics as possible to meet the assumptions of the Bayesian classifier and improve the accuracy and reliability of application classification.

\subsection{Bayes Updating Model}
After the classifier model is established, with the growth in application time, new network applications increase and result in a decline in classification accuracy in the original traffic classification model. Accordingly, this paper proposes a new traffic update approach after an interval and adopts the new network dataset to update the original network classification model to improve the classification accuracy and stability of the original model. As mentioned earlier, the Bayesian classifier model assumes that data attributes are independent and comply with the Gaussian distribution; thus, for a given training data set $D={x_1, x_2, ..... x_n}$, it can be deduced that
\begin{equation}
f(x|\mu,\sigma^2)=\frac{1}{\sqrt{2\Pi \sigma^2}}exp[-\frac{1}{2}(\frac{x-\mu}{\sigma})^2]
\end{equation}
Each data is independent; hence,
\begin{equation}
\begin{aligned}
p(D|\mu,\sigma^2)=\prod_{i=1}^n p(x_i|\mu,\sigma^2)\\=(2\Pi \sigma^2)^{-\frac{n}{2}}exp[-\frac{1}{2\sigma^2} \sum_{i=1}^n(x_i-\mu)^2]
\end{aligned}
\end{equation}
Known dataset $D={x_1,x_2,...,x_n}$, so the following formula is obtained:
\begin{equation}
p(x_1,x_2,...,x_n|\mu,\sigma^2)\propto \frac{1}{\sigma^n} exp[-\frac{1}{2\sigma^2} \sum_{i=1}^n(x_i-\mu)^2]
\end{equation}
\begin{equation}
p(D|\mu,\sigma^2)=(\frac{1}{(2\Pi)^{-n/2}} \sigma^2)^{-\frac{n}{2}}exp[-\frac{1}{2\sigma^2}[ns^2+(x_i-\mu)^2]]
\end{equation}
Using the derivation of conjugate Bayesian theory based on the Gaussian distribution, the posterior probability distribution is
\\ $p(\mu,\sigma^2)\sim NIG(p_n,q_n;,a_n,b_n)$\\
$p_n=\frac{k_0}{k_0+n}p_0+\frac{n}{k_0+n}\overline{x}$\\
$q_n=\frac{1}{v_0+n}$\\
$k_n=k_0+n$\\
$v_n=v_0+n$\\
$a_n=a_0+n/2$\\
$b_n=b_0+1/2[p_0^2 q_0^{-1}+\sum_i(x_i^2-p_n^2q_n^{-1})]$\\
which can be further obtained as follows:
\begin{equation}
\mu|\sigma^2,D\sim N(p_n,\sigma^2 q_n)=N[\frac{\frac{k_0}{\sigma^2}\mu_0+\frac{n}{\sigma^2}x\overline{x}}{\frac{k_0}{\sigma^2}+\frac{n}{\sigma^2}},\frac{1}{\frac{k_0}{\sigma^2}+\frac{n}{\sigma^2}}]
\end{equation}
\begin{equation}
\sigma^2|D\sim IG(a_n,b_n)
\end{equation}
where NIG denotes a normal inverse Gaussian (NIG) distribution, and IG represents an inverse Gaussian distribution. Taking into account the new dataset, the probability distribution of the statistical model from Bayesian updating theory leverages new training data to update the traffic classification model and improve flow model accuracy. These changes reduce the system overhead and demonstrate a corresponding increase in overall performance.\\
Specific algorithm steps:\\
1. Sample network traffic training process: for each type i, the training sample is used to calculate the number of each type contained in the flow along with the mean and variance of each type.
\begin{equation}
\widehat{\mu_i}=\frac{1}{n_i}\sum_{x_j \in c_j}x_j
\end{equation}
\begin{equation}
\widehat{\sum_i}=\frac{1}{n_i-1}\sum_{x_j \in c_j}(x_j-\widehat{\mu_i})(x_j-\widehat{\mu_i})'
\end{equation}\\
2, Flow classification process:\\
 for each application category i, the measure feature x is calculated; then, the updated mean and variance of each type are obtained according to Formulas 8-9. The simplified Bayes equation is shown in formula 12:
\begin{equation}
h_i(x)=n_i|\widehat{\sum_i}|^{-1/2}exp{-(x-\widehat{\mu_i})'\widehat{\sum_i}^{-1}(x-\widehat{\mu_i})/2}
\end{equation}
where $h_i(x)$ is the Bayes probability of each application type. If $h_i(x)$ is the largest, then i will be considered as the number of application protocols. The algorithm pseudo-code is shown in Algorithm 1:
\begin{algorithm}
\LinesNumbered
\caption{Bayes updating algorithm}
\KwIn{Flow record}
\KwOut{$h_m$}
  Get $\overrightarrow{F}$ from flow record\\
  Update value of every metric\\
  For(1 to j)\\
 \If {$f\in \overrightarrow{F}$}
  {Get mean of each metric from $\overrightarrow{F}$\\
  Get deviation of each metric from $\overrightarrow{F}$}
  \Else{
  goto exit}
 \If{$D_{new}!=null$}{
  for every flow which has set Flow\\
 {$\overrightarrow{x_i}=Get_{metric}(M_{si}'')$\\
   $mean_{update}=Get_{mean}(M_{si}'')$\\
   $dev_{update}=Get_{dev}(M_{si}'')$\\
   $h_i=NN_{S_i}(\overrightarrow{x_i})$\\
   $h_m=Max({h_i})$\\
   return $h_m$\\}}
\end{algorithm}
Nowadays most current research focuses on data collected for full packets in traffic identification fields; more information packets generally lead to more accurate traffic identification and classification results. However, this method carries a large computational cost and high calculation complexity, making it infeasible for online identification. This study thus considers the inherent and extended NETFLOW flow statistical characteristics as flow features to identify traffic, which can reduce pressure from heavy traffic and improve identification accuracy to realize actual online traffic identification. As such, NETFLOW and extended NETFLOW flow records are taken as the study object; the proposed NFI model is shown in Figure 1. NETFLOW, extended flow records, and application type are defined below followed by an introduction to the proposed model.
\begin{df}
NETFLOW flow records and extended flow record;\\
$X=\{x_1,x_2,x_3,...,x_t\}$
\end{df}
\begin{df}
Application Type:objective result of identification;\\
$Y=F(X)=\{y_1,y_2,y_3,...,y_t\}$
\end{df}
Function parameters can be determined by the training sample data; the classifier is a function F(X) itself.
\begin{figure*}
\centering

\includegraphics[width=0.7\textwidth]{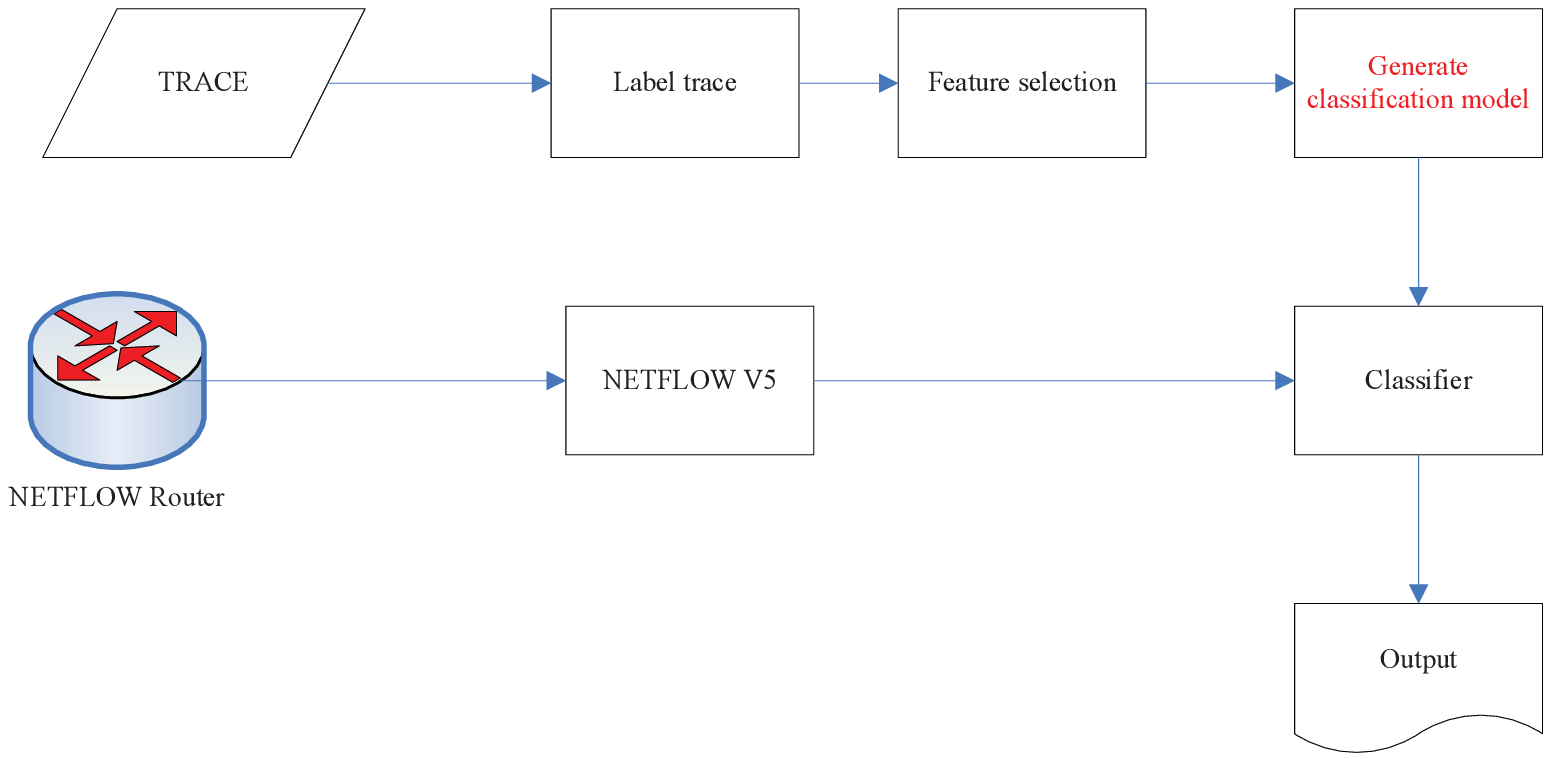}
\centerline{\footnotesize\begin{tabular}{c} Fig.\ 1.\ NETFLOW traffic identification process-based machine learning
\end{tabular}}
\label{fig:2}       
\end{figure*}

\begin{algorithm}
\LinesNumbered
\caption{NFI algorithm}
\KwIn{Full packets from network}
\KwOut{Classification result}
collect the full packets from network\\
Get $\overrightarrow{F}$ by grouping flow\\
Update value of every metric\\
\For{1 to j}{
\If{$f\in \overrightarrow{F}$}{
$\overrightarrow{F_{new}}$=FCBF($\overrightarrow{F}$)\\
Get new metric from $\overrightarrow{F}$ by feature selection algorithm\\
Bayes-updating classifer=Bayes-updating training($\overrightarrow{F_{new}}$)\\
results=Bayes-updating classifer($\overrightarrow{F_{new}}$)\\
return results\\}
}
\end{algorithm}
Figure 1 depicts the online traffic identification model based on extended NETFLOW flow. The model is divided into four phases: data collection, feature selection, model of traffic classification (i.e., training process), and traffic classification. Data are gathered from full packets, using DPI tools (l7filter) to label traffic. The feature selection process is intended to construct the NETFLOW standard and extended flow metrics(see experimental section for details). The traffic classification model establishes the appropriate classifier based on machine learning algorithms. In the NFI algorithm shown in Algorithm 2, the classifier-building process can be completed using the Bayes-updating training method.  Traffic classification serves to identify classification by establishing a classifier based on the NETFLOW V5 format and extended NETFLOW records, which is process completed using the Bayes-updating classifier. Finally, the corresponding classification results are obtained.
 \begin{table*}
\label{tab:1}       
\begin{center}
\begin{tabular}{lll}
\multicolumn{2}{c}{\bf Table 1.\ Predominant feature used to describe}\\
\hline\noalign{\smallskip}
\centering
Feature & Feature Discription \\
\noalign{\smallskip}\hline\noalign{\smallskip}
lport& low port number\\
hport& high port number\\
duration & Flow duration\\
Transproto & traffic transport protocol used (TCP / UDP)\\
TCPflags1 & TCP header flag, or (OR), transport layer protocol is UDP, the feature is 0\\
TCPflags2 & TCP header flag, or (OR), transport layer protocol is UDP, the feature is 0\\
pps & Packets/duration\\
bps & bytes/duration\\
Mean packets arrived time & duration/packets\\
Bidirectional Packets  ratio & Forward packets/ backward packets\\
Bidirectional Bytes ratio & Forward bytes/ backward bytes\\
Bidirectional Packet length ratio & Bidirectional packets length ratio\\
Bidirectional packets & Forward packets + backward packets\\
Bidirectional bytes & Forward bytes + backward bytes\\
tos & Bidirectional TOS OR from NETFLOW\\
Mean packet length & Bidirectional bytes/Bidirectional packets\\
\noalign{\smallskip}\hline
\end{tabular}
\end{center}
\end{table*}

\subsection{Analysis of netflow sampling on flow behavior}
In this study, netflow sampling was applied to traffic identification to improve the identification efficiency and to influence flow behavior along with various changes in the classification accuracy. This section addresses sources of inaccuracy due to excessive netflow features. Only three features (flow length, flow size, and flow duration) are considered research objects.
\begin{df}
flow length:let flow length be l, denoting the total number of packets in a flow.
\end{df}
\begin{df}
Flow size: let flow size be s, $s_i$  is the size of each individual packet.Flow size s is denoted as
\begin{equation}
s=\sum^n_{i=1}s_i
\end{equation}
where n represents the total number of packets in the flow.
\end{df}
\begin{df}
Flow duration:let flow duration be $fd,fd=t_n-t_1$,where $t_1$ and $t_n$ are the timestamps of the first and last packets in
the original flow, respectively.
\end{df}
Here, We present a theoretical analysis of the error in the feature estimation using sampling theory. Suppose the sampling ratio is p,after packets sampling, let the flow length, flow size and flow duration as $\widehat{l}$, $\widehat{s}$ and $\widehat{fd}$.Suppose sampled $l_sampled$ packets from flow length as $l$,then $\widehat{l}=\frac{l_{sampled}}{p}$,mean value of $\widehat{l}$ is denoted as $E(\widehat{l})$, which is calculated by
\begin{equation}
E(\widehat{l})=E(\frac{l_{sampled}}{p})=\frac{1}{p}E(l_{sampled})=\frac{1}{p}lp=l
\end{equation}
where $\widehat{l}$ is represented by unbiased estimate of l,then variance of $\widehat{l}$ is denoted as:
then the variance for the relative error of $\widehat{l}$ is
\begin{equation}
\begin{aligned}
var(1-\frac{\widehat{l}}{l})=var(\frac{\widehat{l}}{l})=\frac{1}{l^2}var(\widehat{l})=\frac{1}{l^2}lp(1-p)\\=\frac{1}{p}l(1-p)
\end{aligned}
\end{equation}
The analysis indicates that the short flow has a larger error than the long flow, when the sampling ratio p for flow size is certain, because $s=\sum^n_{i=1}s_i$, then $\widehat{s}=\sum^n_{i=1}t_i\frac{s_i}{p}$, where $t_i\in{0,1}$ are Bernoulli distributed random variables.
\begin{equation}
\begin{aligned}
E(\widehat{s})=E(\sum^n_{i=1}t_i\frac{s_i}{p})=\frac{1}{p}E(\sum^n_{i=1}t_i s_i)=\frac{1}{p}\sum^n_{i=1}E(t_i s_i)\\=\frac{1}{p}\sum^n_{i=1}s_i E(t_i)=\frac{1}{p}\sum^n_{i=1}s_i p
=\frac{1}{p}p\sum^n_{i=1}s_i=s
\end{aligned}
\end{equation}
Thus, $\widehat{s}$ is unbiased estimate of s, so the variance of $\widehat{s}$ is
\begin{equation}
\begin{aligned}
var(\widehat{s})=var(\sum^n_{i=1}t_i\frac{s_i}{p})=\frac{1}{p^2} var(\sum^n_{i=1}t_i s_i)\\=\frac{1}{p^2}\sum^n_{i=1}var(t_i s_i)\\=\frac{1}{p^2}\sum^n_{i=1}{s_i}^2 var(t_i)\\
=\frac{1}{p^2}\sum^n_{i=1}{s_i}^2 p(1-p)\\=\frac{p-1}{p}\sum^n_{i=1}{s_i}^2
\end{aligned}
\end{equation}
and the variance for the relative error of $\widehat{s}$ is
\begin{equation}
\begin{aligned}
var(1-\frac{\widehat{s}}{s})=var(\frac{\widehat{s}}{s})=\frac{1}{s^2}var(\widehat{s})\\=\frac{1}{s^2}\frac{1-p}{p}\sum^n_{i=1} {s_i}^2\\=\frac{p-1}{p}\frac{\sum^n_{i=1} {s_i}^2}{(\sum^n_{i=1} {s_i})^2}
\end{aligned}
\end{equation}
for flow duration, suppose flow duration as $fd=t_n-t_i$, where $t_i$ and $t_n$ are the timestamps of the first and last packets of the original flow. In sampling environment, $\widehat{fd}$ is estimate of fd,$\widehat{fd}=t_b-t_a$, where $t_a$ and $t_b$ are the timestamps of the first and last packets of the sampled flow. Mean value of $\widehat{fd}$ is denoted as $E(\widehat{fd})$, which is calculated by
\begin{equation}
\begin{aligned}
E(\widehat{fd})=E(t_b-t_a)=E(t_b)-E(t_a)\\=E(t_n-\sum^n_{i=b}iat_i)-E(t_1+\sum^a_{i=1}iat_i)\\
=(t_n-t_1)-(E(\sum^n_{i=b}iat_i)+E(\sum^a_{i=1}iat_i))
\end{aligned}
\end{equation}
where $iat_i$ represents time interval from ith to (i-1)th packets arrival.The corresponding error expectation and variance are derived by the theoretical analysis on three flow features metric. In this paper, which is defined as Degree of Relative Error(DRE).
\begin{equation}
DRE=\left\{\begin{array}{ll}
E(M_i-\widehat{M_i}),&\mbox{if metric $M_i$ is biased}\\
var(\frac{M_i-\widehat{M_i}}{M_i}),&\mbox{if metric $M_i$ is unbiased}\\
\end{array}\right.
\end{equation}
where $M_i$ is a feature metric in flow,$\widehat{M_i}$ is a flow feature metric in sampling environment,$i$ is metric number.In real network environment, in order to better evaluate the Degree of Relative Error(DRE)of a metric,Average Degree of Relative Error(ADRE) will be introduced in this paper. ADRE is expressed as
\begin{equation}
ADRE=\left\{\begin{array}{ll}
\sum^n_{k=1}E_k(M_i-\widehat{M_i})/n,&\mbox{if metric $M_i$ is biased}\\
\sum^n_{k=1}var_k(\frac{M_i-\widehat{M_i}}{M_i})/n,&\mbox{if metric $M_i$ is unbiased}\\
\end{array}\right.
\end{equation}
where $n$ represents flow number,$E_k(M_i-\widehat{M_i})$ means relative error of ith metric $M_i$ in flow's num k, when metric $M_i$ is biased, $var_k(\frac{M_i-\widehat{M_i}}{M_i})$ represents relative error of ith metric $M_i$ in flow's num k, when metric $M_i$ is unbiased.
\begin{table}
{\tabcolsep=2.5pt \footnotesize
\label{tab:1}       
\begin{center}
\begin{tabular}{lllllll}
\multicolumn{5}{c}{\bf Table 2.\ ADRE of typical feature metrics}\\
\hline\noalign{\smallskip}
\centering
 & p=1:128 & p=1:256& p=1:512& p=1:1024\\
\noalign{\smallskip}\hline\noalign{\smallskip}
Flow length& 3.72& 6.78& 29.32& 55.36\\
Flow size& 3.89& 7.23& 29.79& 56.24\\
Flow duration & 0.66& 0.74& 0.78& 0.79\\

\noalign{\smallskip}\hline
\end{tabular}
\end{center}
}
\end{table}
in order to analyze ADRE of the typical feature metrics, the SKYPE-SET will be used in this experiment. The experimental results is presented in table 2 listed all relative error, in which the sampling ratio varies from 1:1024 to 1:128, we can note that the relative error of each metric is reducing with sampling ratio increasing. From overall results, flow length and flow size have higher relative error, while flow duration is low. Due to exist much short flow, according to above theory, in the sampling environment, when sampling ratio is concert, then its ADRE is larger, when p=1:1024, ADRE of metric flow length can reach 55.36 and flow size is 56.24.

\section{Experiment Results and Analysis}

\subsection{Experimental evaluation}
This paper uses the routine evaluation standard to verify the effectiveness of the proposed classification algorithm. The effectiveness of the current flow identification algorithm includes the following four evaluation criteria: TPR, FPR, TNR, and FNR. Relevant concepts are as follows:\\
TP (true positive): the flows of application A are correctly classified as A, which is a correct classification result.\\
FP (false positive): the flows outside of A are misclassified as A (e.g., a non-P2P flow is misclassified as a P2P flow). FPs produce false warnings in the classification system. \\
TN(true negative):the flows of application A are not correctly classified as A, which is a correct classification result.\\
FN (false negative):  the flows in A are misclassified as belonging to some other category (e.g., a true P2P flow is not identified as P2P); FNs will result in a loss of classification accuracy.\\
TPR:true positive rate
\begin{equation}
 TPR=\frac{TP}{TP+FN}
\end{equation}
FPR:false positive rate
\begin{equation}
FPR=\frac{FP}{FP+TN}
\end{equation}
TNR:true negative rate
\begin{equation}
TNR=\frac{TN}{TN+FP}
\end{equation}
FNR:false negative rate
\begin{equation}
FNR=\frac{FN}{FN+TP}
\end{equation}
Precision,Recall,Overall Accuracy and F-measure:
\begin{equation}
Precision=\frac{TP}{TP+FP}
\end{equation}
\begin{equation}
Recall=\frac{TP}{TP+FN}=TPR
\end{equation}
\begin{equation}
Overall Accuracy(OA)=\frac{TP+TN}{TP+FP+TN+FN}
\end{equation}
\begin{equation}
F-measure=\frac{2*Precision*Recall}{Precision+Recall}
\end{equation}
\subsection{Experiment and dataset}

Four different traces were used for accuracy. The first and second traces are called Trace 1 and Trace 2 , containing Skype traffic captured on the access link of Politecnico di Torino\cite{Han2005Borderline}. The set of users in such a network typically consists of students, faculty, and administrative staff. The measurement campaign duration was 96 hours in May–June 2006. Trace 1 only contains end-to end Skype voice and video calls, whereas Trace 2 only contains outbound Skype calls. Trace 1 contains 40M packets, and Trace 2 contains 3M packets. The third trace, Trace 3 \cite{Japkowicz2002The}, is a synthetic trace captured in the laboratory at Universidad Autonomade Madrid in August 2008. The trace contains 22M packets of P2P traffic from several applications, such as Emule and Bittorrent. Traces 1 and 2 could only be used to estimate the false negatives rate, as these traces only contain Skype traffic; however, Trace 3 could be used to estimate the false positives rate as the trace does not contain Skype traffic. The false negative rate was below 1\% in Trace 1 and approximately 6\% in Trace 2. Trace 3 exhibited a false negative rate equal to zero. The obtained accuracy results are similar to those reported in previous works using Traces 1 and 2.  \cite{He2009Learning} found a false negative rate around 6\% (in bytes) in the best-case scenario using only statistical classifiers (i.e., without inspecting packet payload). \cite{Cieslak2012Hellinger}, obtained a false negative rate higher than that in the current study  when only using the na?ve Bayes classifier. The present research connected Skypeness to a server, which reproduces a pcap file using Tcpreplay. This tool allows for transmission of pcap traces at a variable rate. The transmission rate varied during tests (100, 250, 500, 750, and 1000 Mbps). A limitation was identified in the Tcpreplay throughput to 1 Gbps (i.e., the pcap file could not be sent faster than 1 Gbps despite using 10 Gbps NICs). In this experimental setup, only one reception queue and one traffic classifier instance were running in the server, requiring only two cores: one for receiving packets and one for detecting Skype flows. In terms of NUMA affinity, the CPU affinity of the reception queue to NUMA node 1 and the CPU affinity of the Skype detector to NUMA node 4 (the worst case in terms of distance). The last trace, Trace 4 \cite{Brefeld2005AUC}, was captured from a 3G access network of a Spanish provider. The full trace contains traffic from residential households and small businesses. The trace also contains 70M packets corresponding to 12M TCP/UDP flows captured over 18 hours in June 2009. For each speed step, the bit rate, packet rate, maximum number of expired flows (and consequently analyzed) per second, and packet loss rate in the whole trace were observed. No packet loss occurred. Again, these results were obtained using only two cores: one for receiving packets to store them in memory, and one for traffic classification. By using the technique proposed in \cite{Sun2010On},which assigns one reception queue per socket, the present study set up to 16 reception queues and 16 detection processes to determine whether this technique would allow performance gains of 16x, thus enabling 10 Gbps of Skype traffic classification in a general-purpose server. Some offline experiments were conducted to address this issue. These experiments used a modified version of Skypeness software that obtains traffic from a local pcap trace instead of opening a socket for frame reception. The theoretical read/write throughput of the chosen DDR3 memory was 170.6 Gbps, much larger than the bandwidth of an Internet backbone link. To compute the hypothetical bandwidth that Skypeness can handle, the program was executed 10 times, and execution times were obtained using Trace 4 as a source. This methodology was repeated by incrementing the number of parallel instances of the Skypeness process and obtaining the corresponding execution times. The four trace data are shown in Table 3.\\
\begin{table*}
\center
\begin{tabular}{llllll}
\multicolumn{5}{c}{\bf Table 3.\ Trace}\\
\hline\noalign{\smallskip}
Trace & start & duration & bandwidth & bps(M) & pps(K) \\
\noalign{\smallskip}\hline\noalign{\smallskip}
Trace1 & May. 10, 2006 14:00 & 96 hour & 1G*2*3 & 3120 & 640\\
Trace2 & June. 20, 2006 15:00 & 96 hour & 1G*2*3 & 3120 & 640\\
Trace3 & Aug. 20, 2008 16:00 & 20 hour & 1G*2*3 & 3120 & 640\\
Trace4 & June. 14, 2009 09:00 & 18 hours & 2.5Gbps*2 & 846.0 & 138.1\\
\noalign{\smallskip}\hline
\end{tabular}
\end{table*}
 \textbf{Skype-SET:}In order to validate the proposed method and analyze the impact factor, Skype-SET was adopted as a dataset. The feature set was selected as shown in Table 1 from \cite{Wang2010Boosting}. Skypeness data were collected, and the method of grouping flow to construct the Skype-SET was adopted prior to generating the dataset. A basic requirement of traffic classification is correct flow type identification. The flow is clarified by the IP five-tuple consisting of a source IP, destination IP, source port, destination port, and protocol type. To focus on the traffic classification process, semantically complete TCP connections were selected to compose the training sets and testing sets, where semantically complete TCP flow is defined as a bidirectional flow for which one can observe the complete connection setup (SYN-ACK) and another complete connection tear-down (FIN-ACK).

\subsection{Analysis of impact of sampling ratio on Skype traffic identification}
Because Netflow flow was obtained from packets sampling at different sampling ratios(1:128, 1:256, 1:512, 1:1024). \cite{dong2013research} have completed a study on the sampling theory. The section will analyze the impact of sampling ratio for Skype traffic identification according to the sampling theory. Packets sampling can change Skype traffic feature due to different sampling ratios. The impact of sampling was examined based on three features as the research objects (i.e.,flow duration, flow size, flow length). Flow size involves bidirectional bytes($\leq$ 10000bytes), flow length involves bidirectional packets($\leq$ 10000), and the selected flow duration was 3 hours. Experimental data adopted the trace above in which the group flow and computed three features were calculated. The study first analyzed the flow behavior from full packets (no sampling). Experimental results are shown in Fig. 2. Additionally, a different sampling ratio was used to study the impact of sampling on flow behavior; detailed experimental results are shown in Fig. 3-5.
\begin{figure*}[htb]
\centering
\subfigure[Distribution of flow length] {\includegraphics[height=1.5in,width=2in]{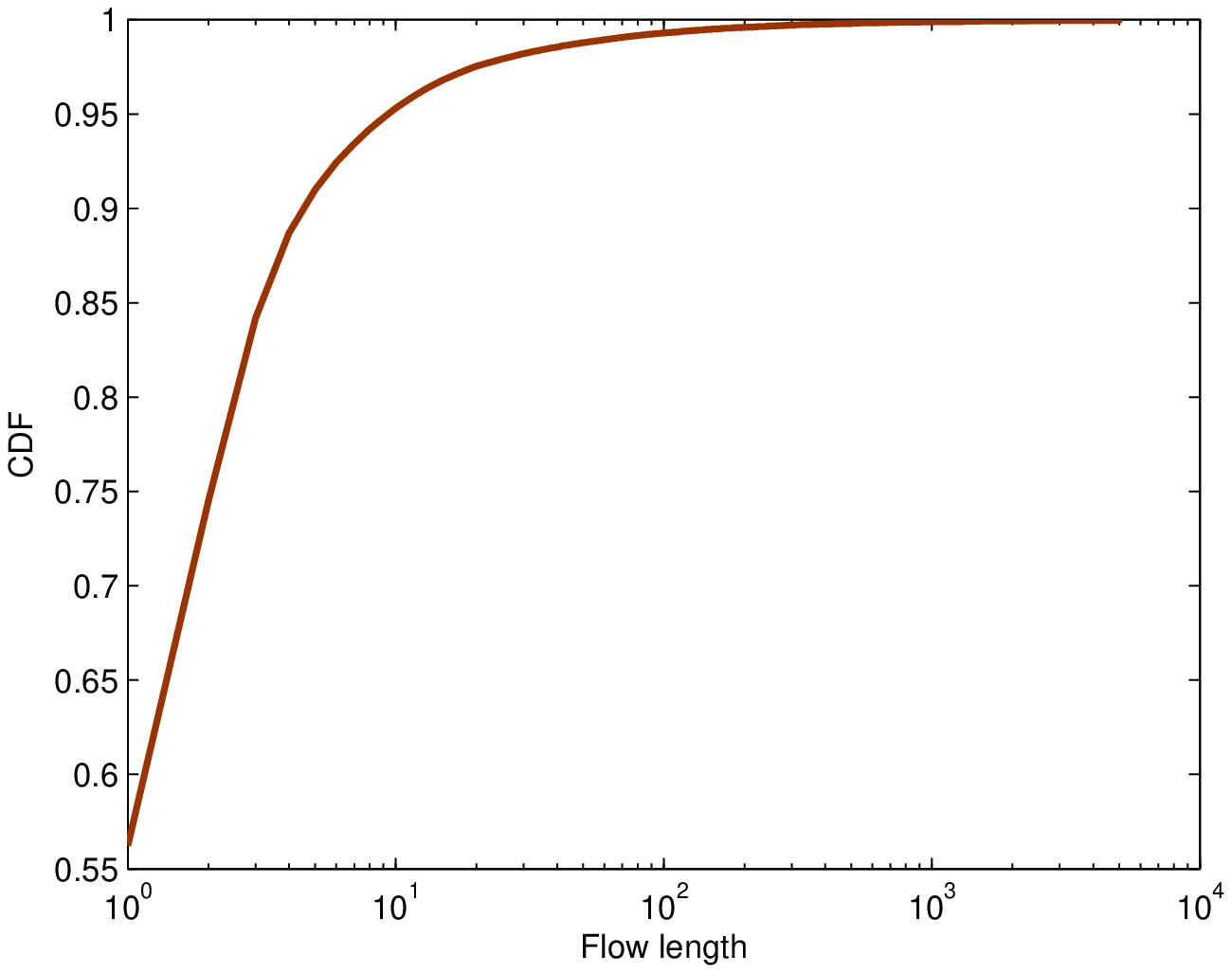}}
\subfigure[Distribution of flow size] {\includegraphics[height=1.5in,width=2in]{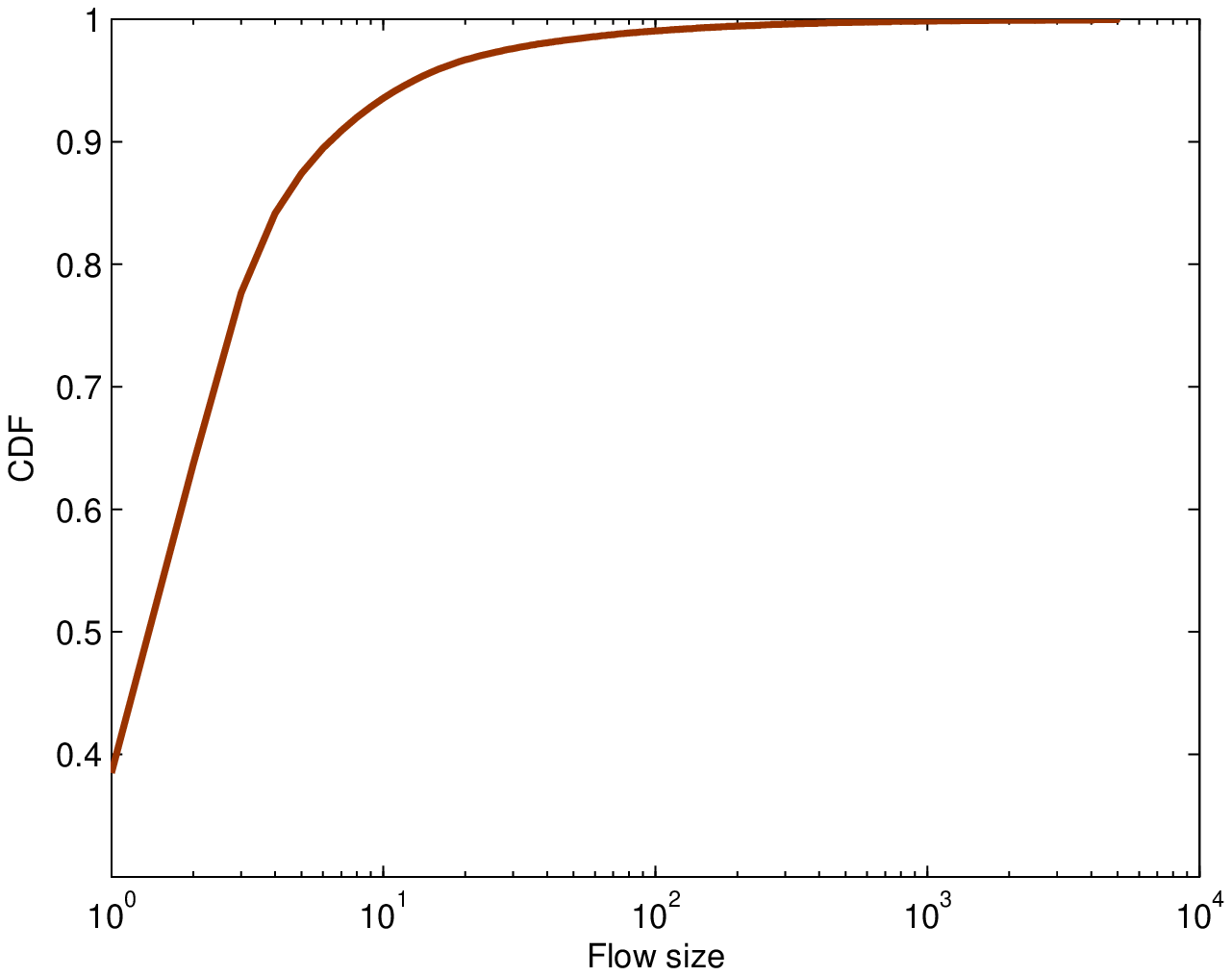}}
\subfigure[Distribution of flow duration] {\includegraphics[height=1.5in,width=2in]{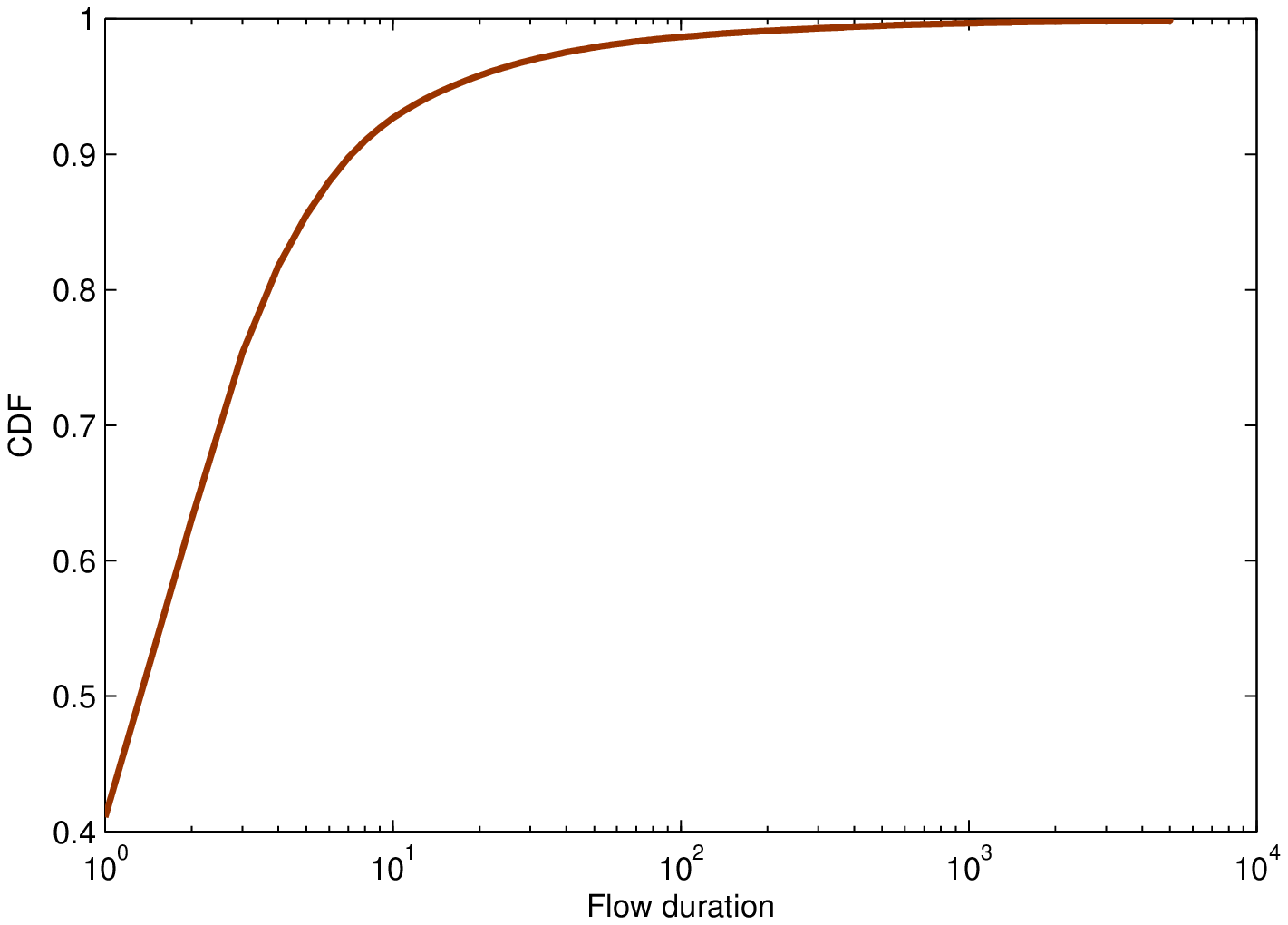}}
 \centerline{\footnotesize\begin{tabular}{c} Fig.\ 2.\ Distribution of typical flow feature
 \end{tabular}}
\label{fig5}
\end{figure*}

 \vskip 4mm

\centerline{\includegraphics[width=0.74\textwidth]{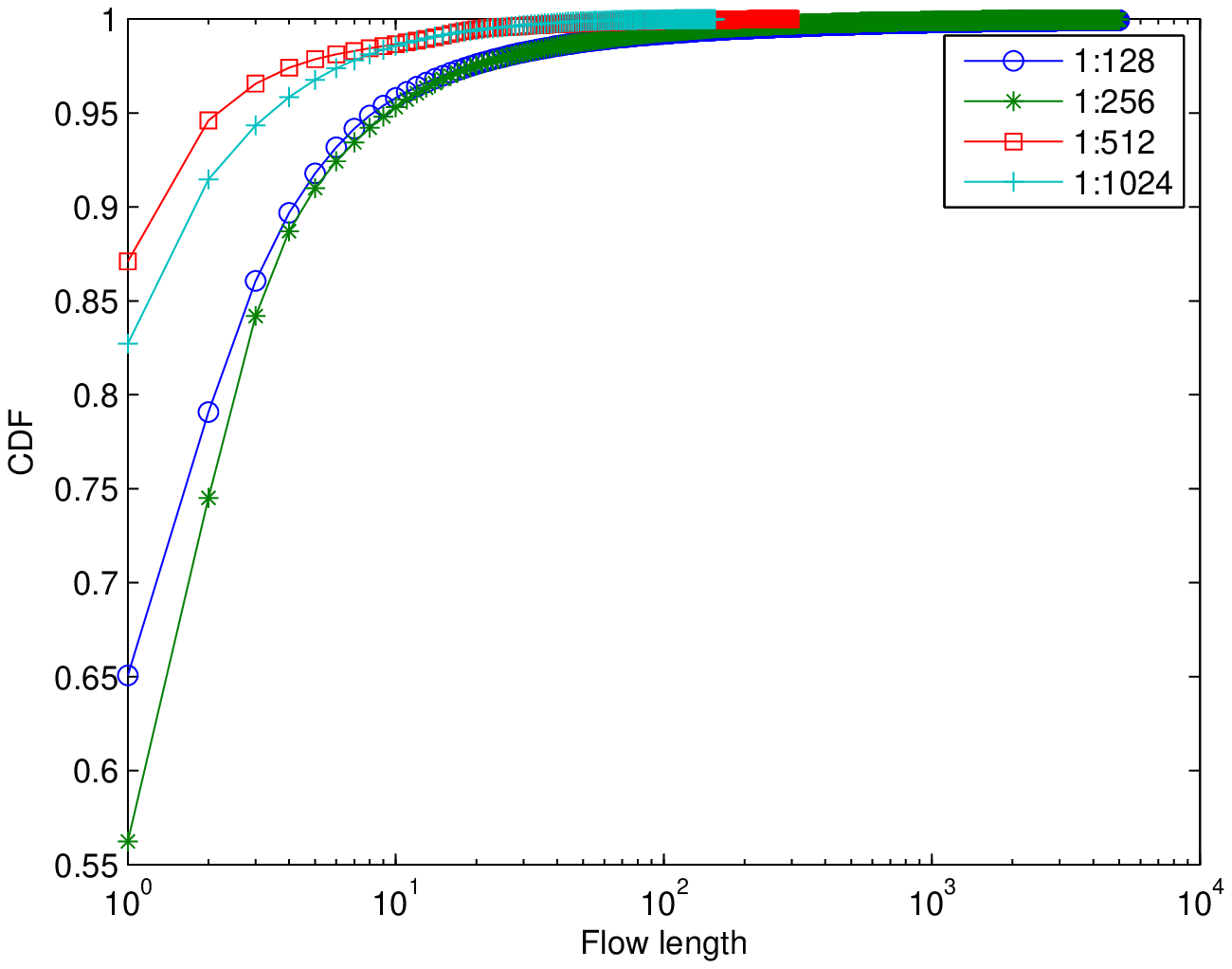} }

\vskip 1mm

\centerline{\footnotesize\begin{tabular}{c} Fig.\ 3.\ CDF of flow length for skype in different sampling ratio
\end{tabular}}

\vskip 0.5\baselineskip


 \vskip 4mm

\centerline{\includegraphics[width=0.74\textwidth]{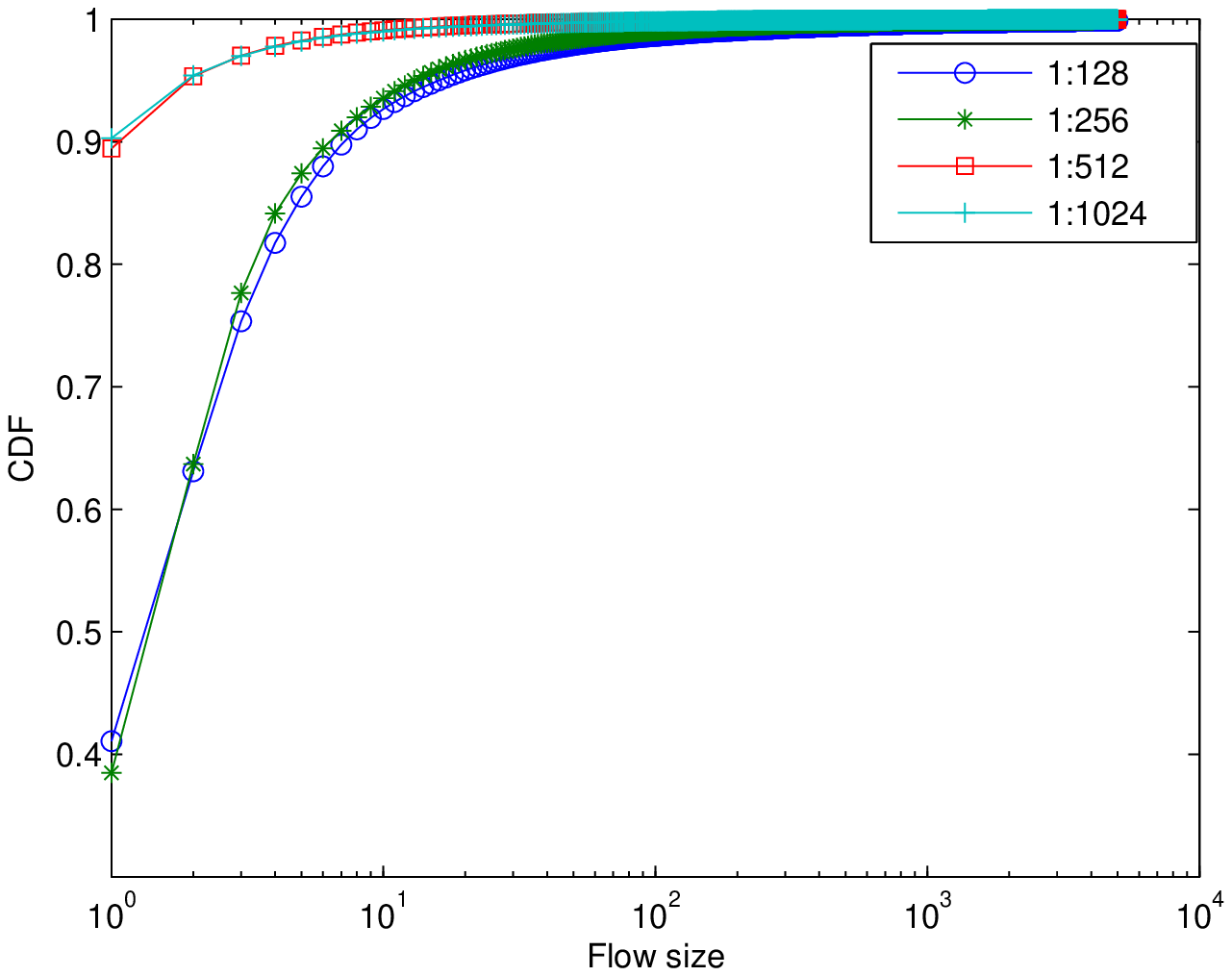} }

\vskip 1mm

\centerline{\footnotesize\begin{tabular}{c} Fig.\ 4.\ CDF of flow size for skype in different sampling ratio
\end{tabular}}

\vskip 0.5\baselineskip
%

 \vskip 4mm

\centerline{\includegraphics[width=0.74\textwidth]{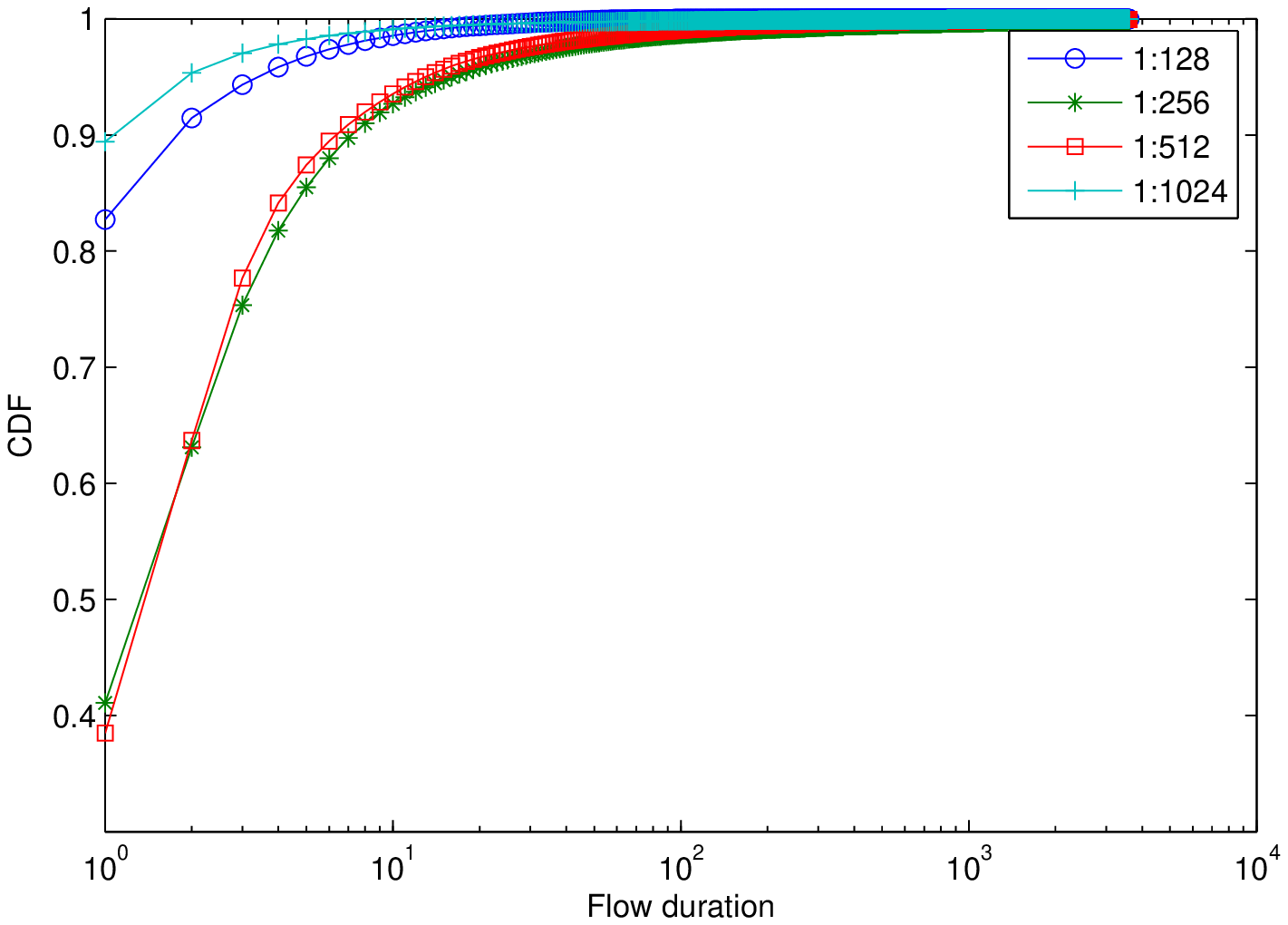} }

\vskip 1mm

\centerline{\footnotesize\begin{tabular}{c} Fig.\ 5.\ CDF of flow duration for skype in different sampling ratio
\end{tabular}}

\vskip 0.5\baselineskip
\begin{figure*}[!htb]
\centering
\subfigure[FPR in different flow length] {\includegraphics[height=2.18in,width=3.18in]{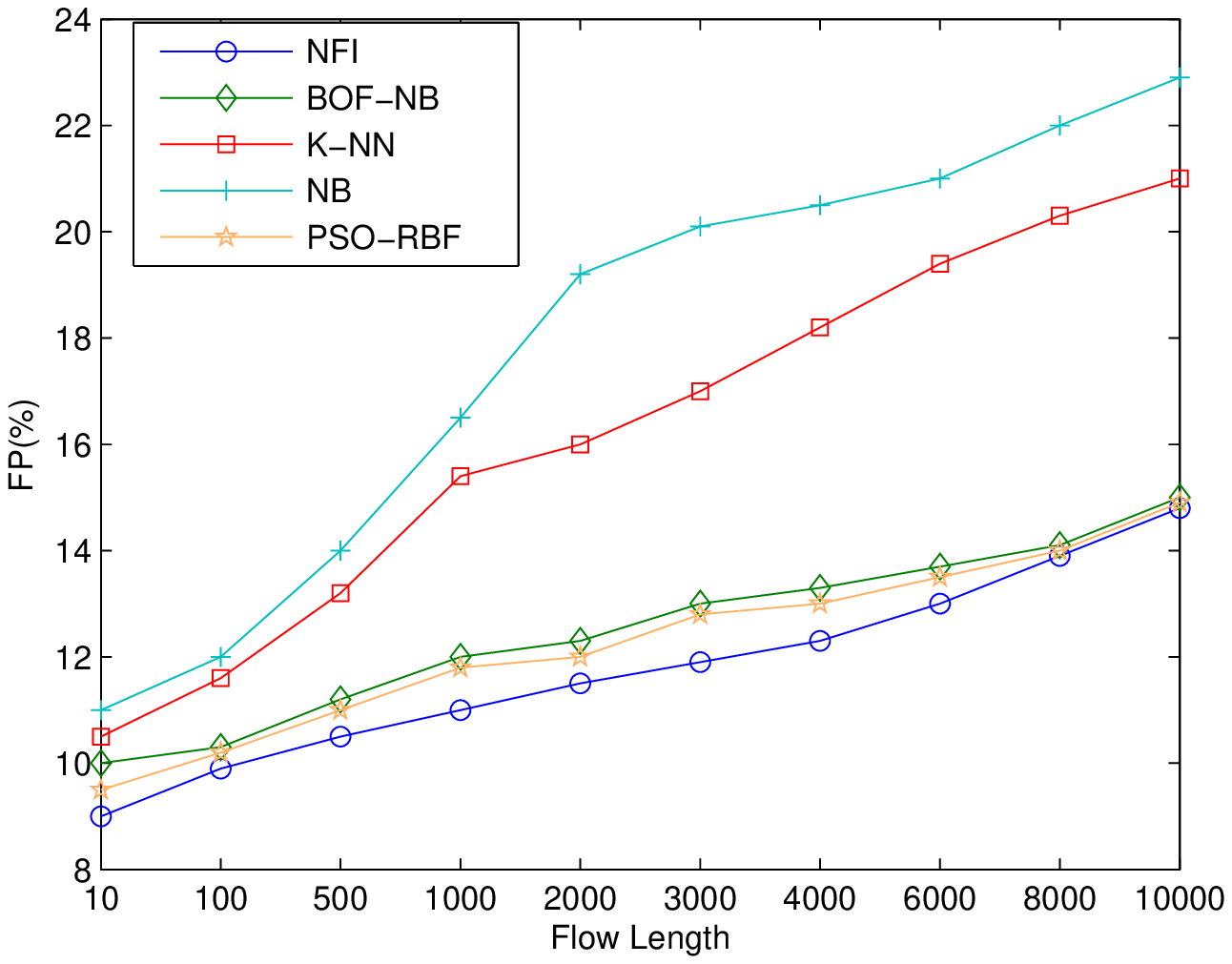}}
\subfigure[FNR in different flow length] {\includegraphics[height=2.18in,width=3.18in]{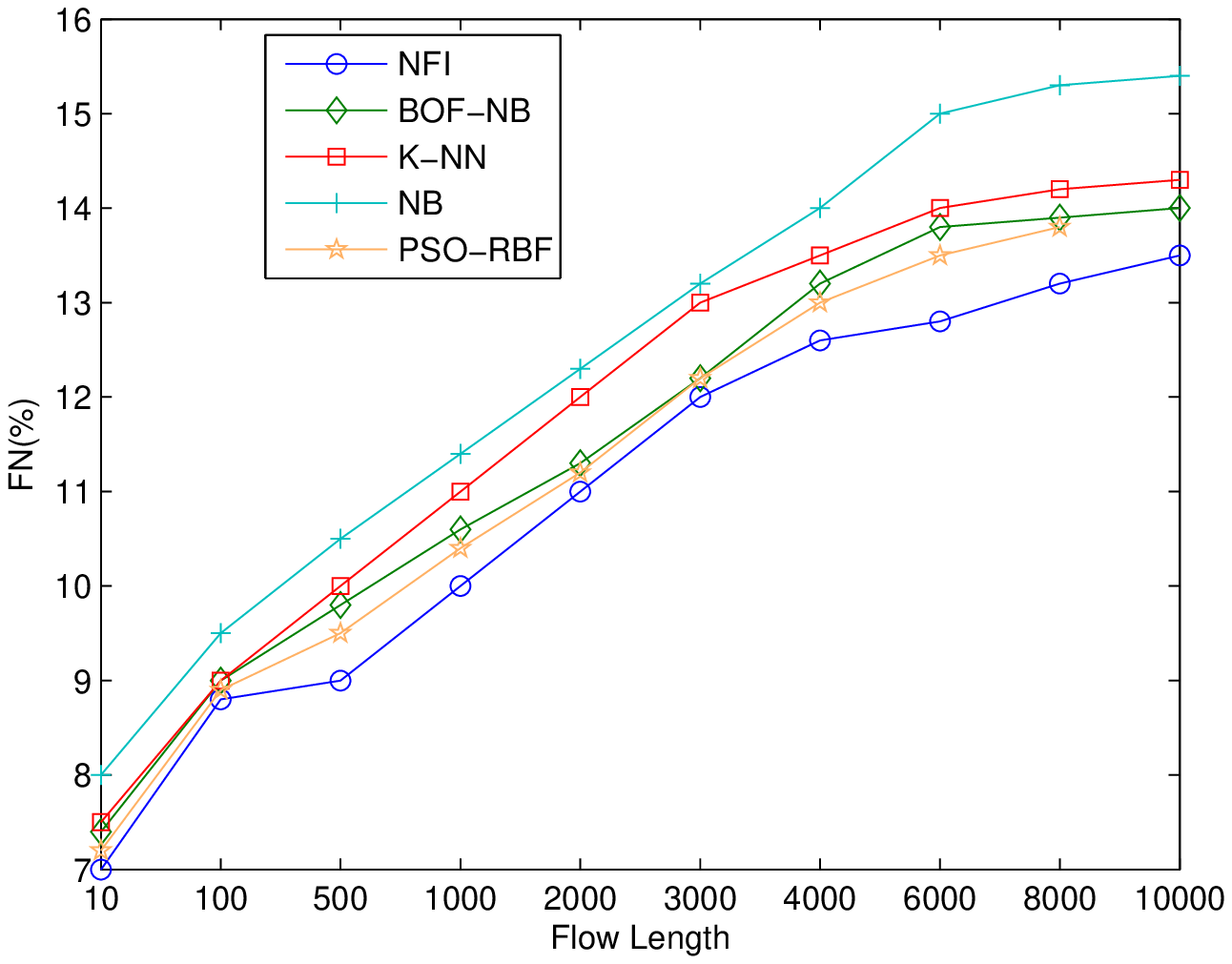}}
\subfigure[TPR in different flow length] {\includegraphics[height=2.18in,width=3.18in]{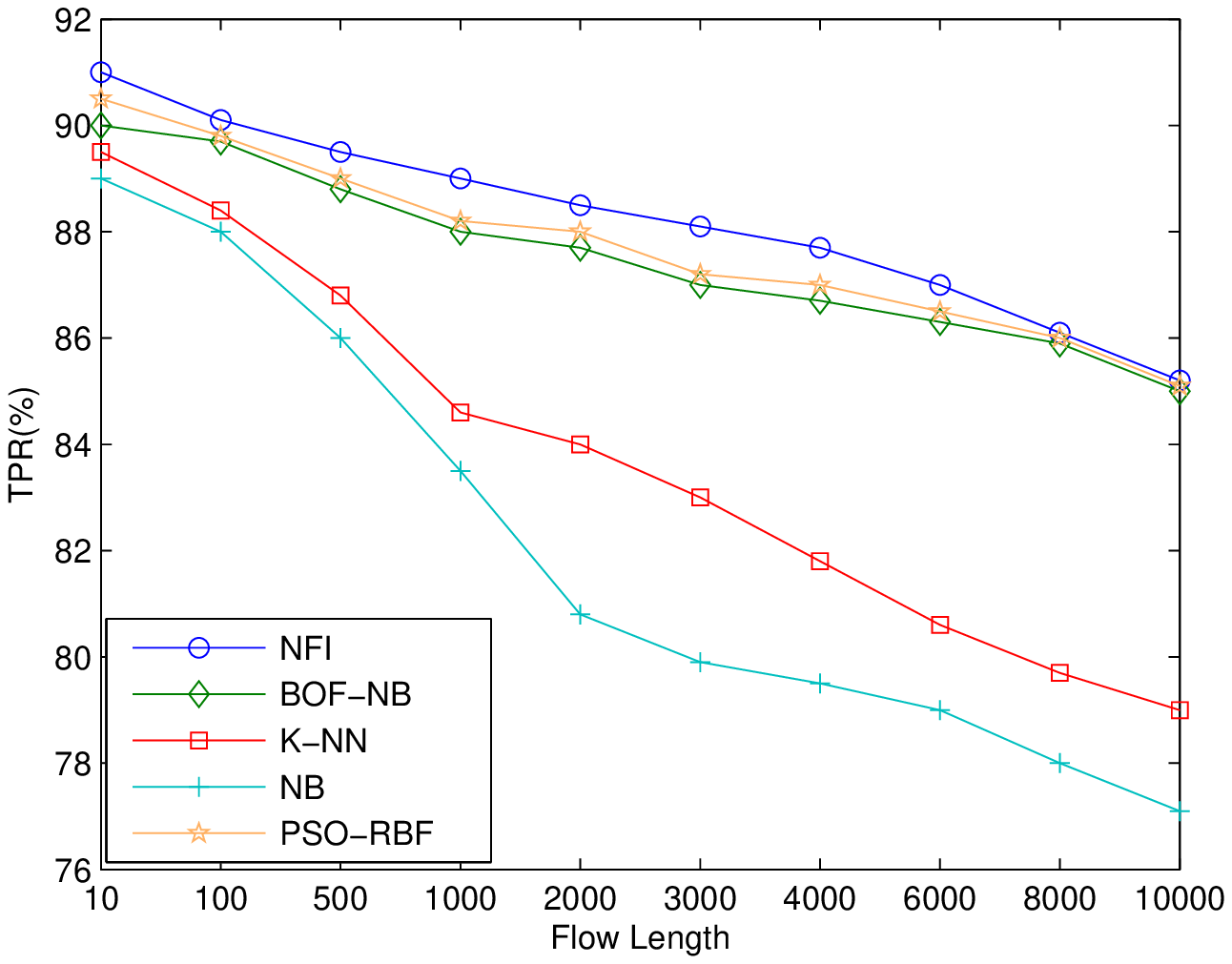}}
\subfigure[TNR in different flow length] {\includegraphics[height=2.18in,width=3.18in]{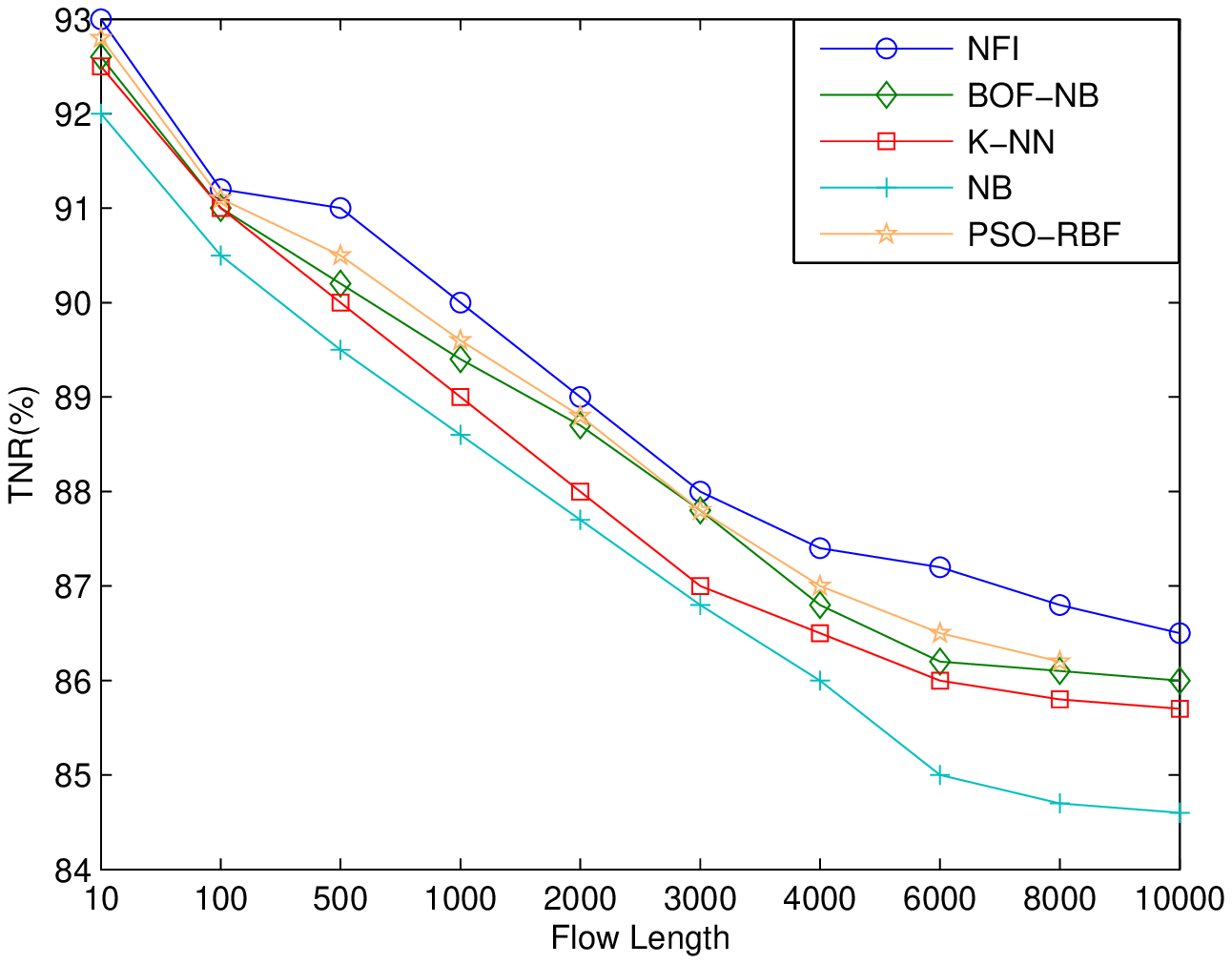}}
\centerline{\footnotesize\begin{tabular}{c} Fig.\ 6.\ Performance evaluation in different flow length
\end{tabular}}
\label{fig5}
\end{figure*}

\begin{figure*}[!htb]
\centering
\subfigure[FPR in different flow size] {\includegraphics[height=2.18in,width=3.18in]{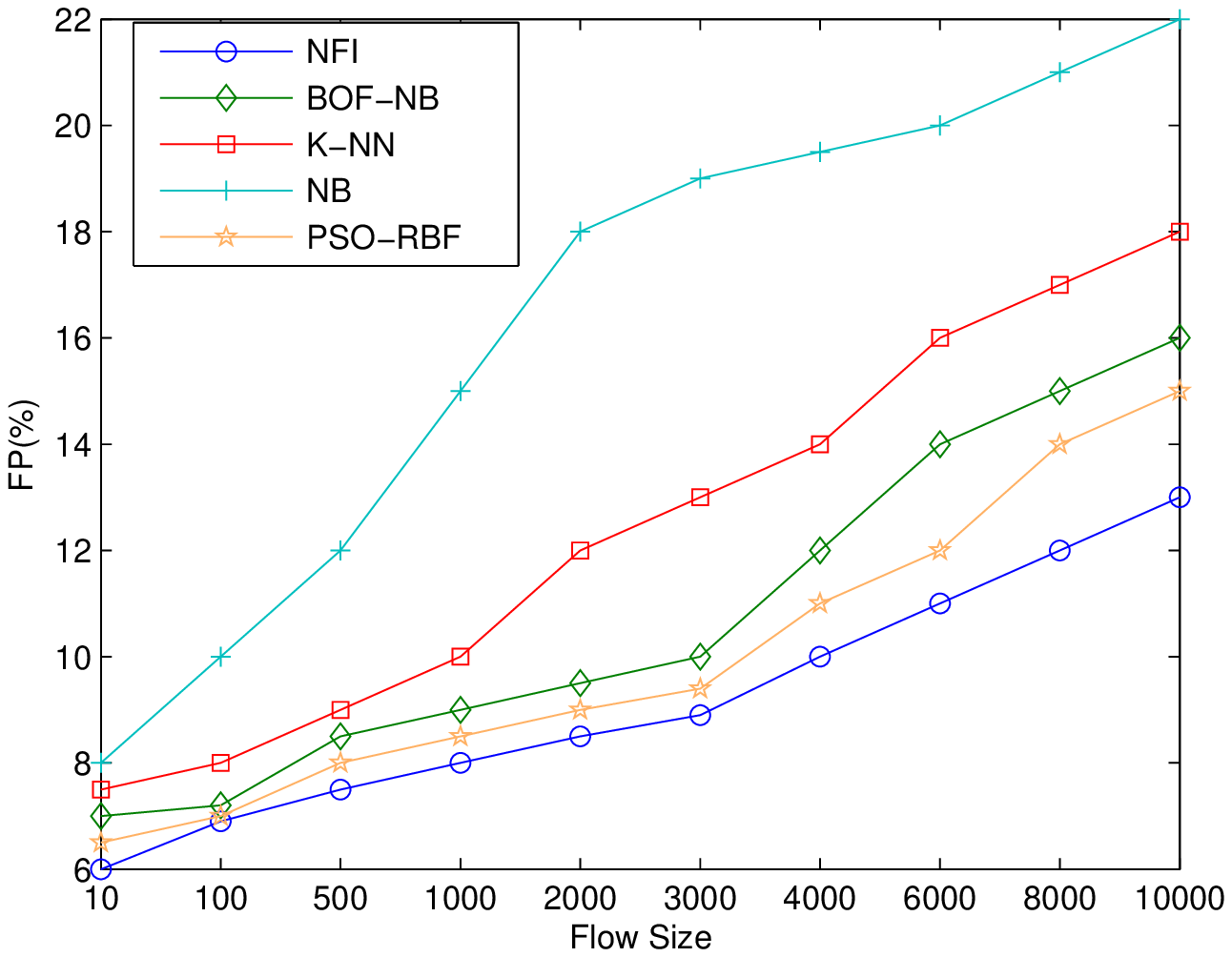}}
\subfigure[FNR in different flow size] {\includegraphics[height=2.18in,width=3.18in]{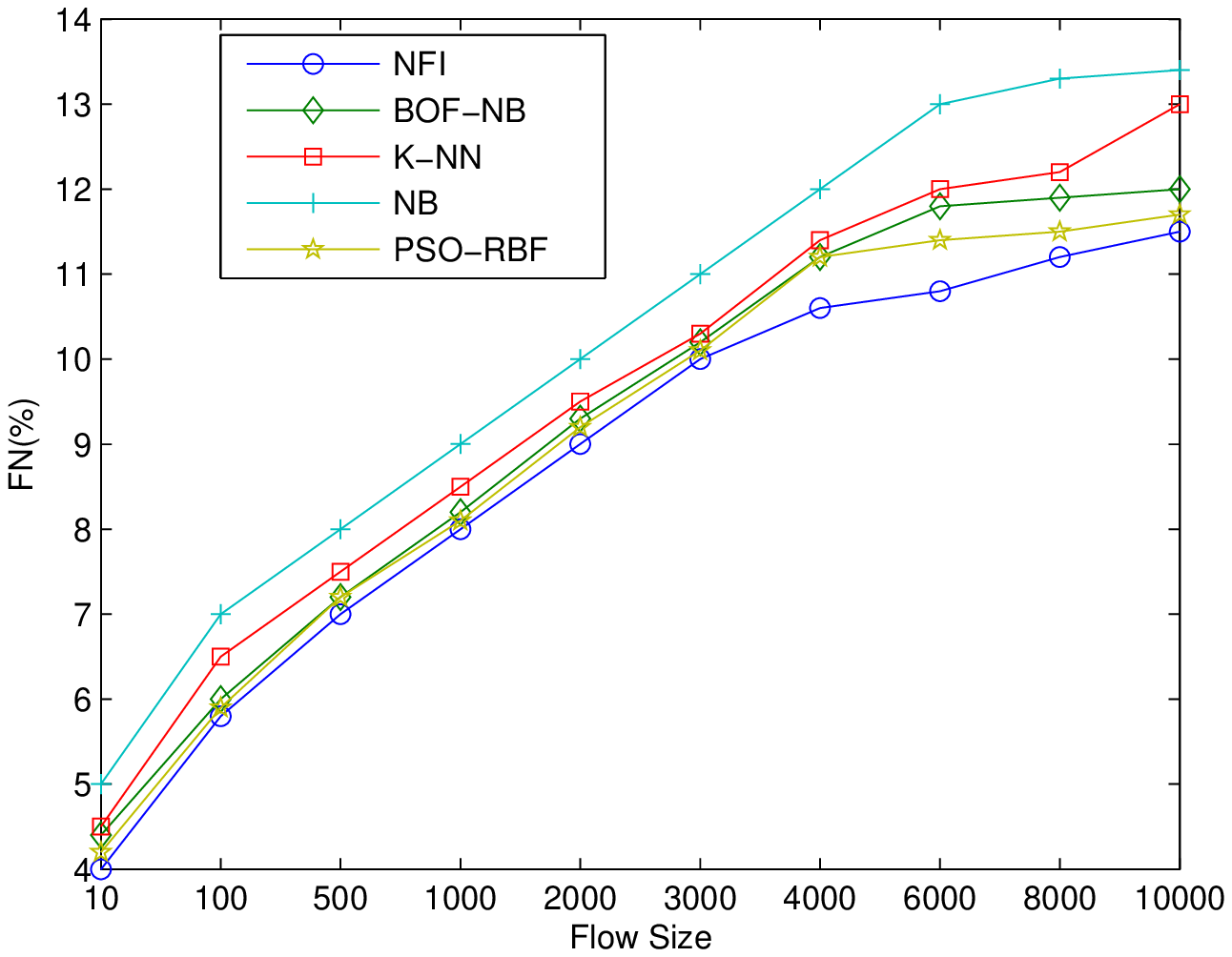}}
\subfigure[TPR in different flow size] {\includegraphics[height=2.18in,width=3.18in]{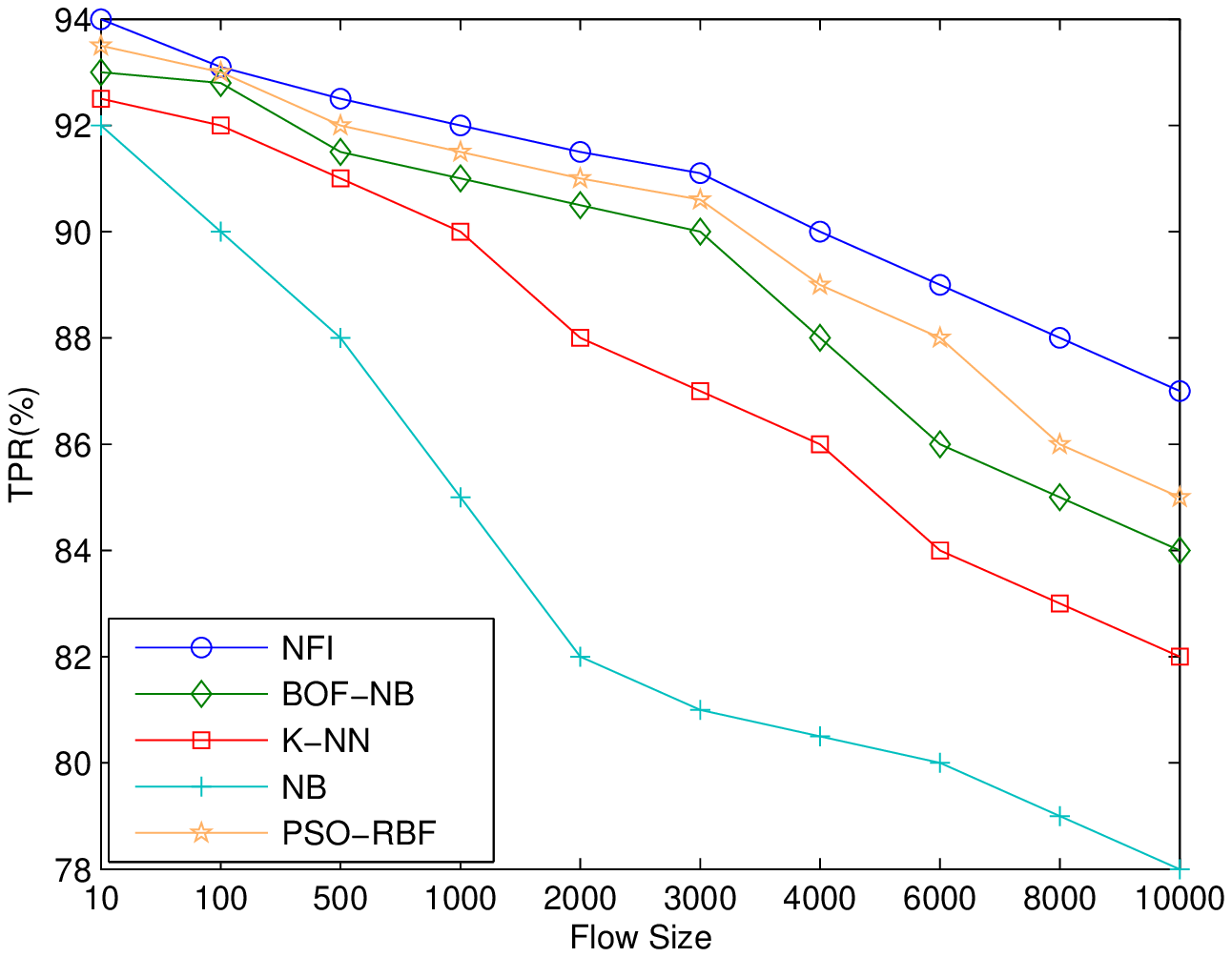}}
\subfigure[TNR in different flow size] {\includegraphics[height=2.18in,width=3.18in]{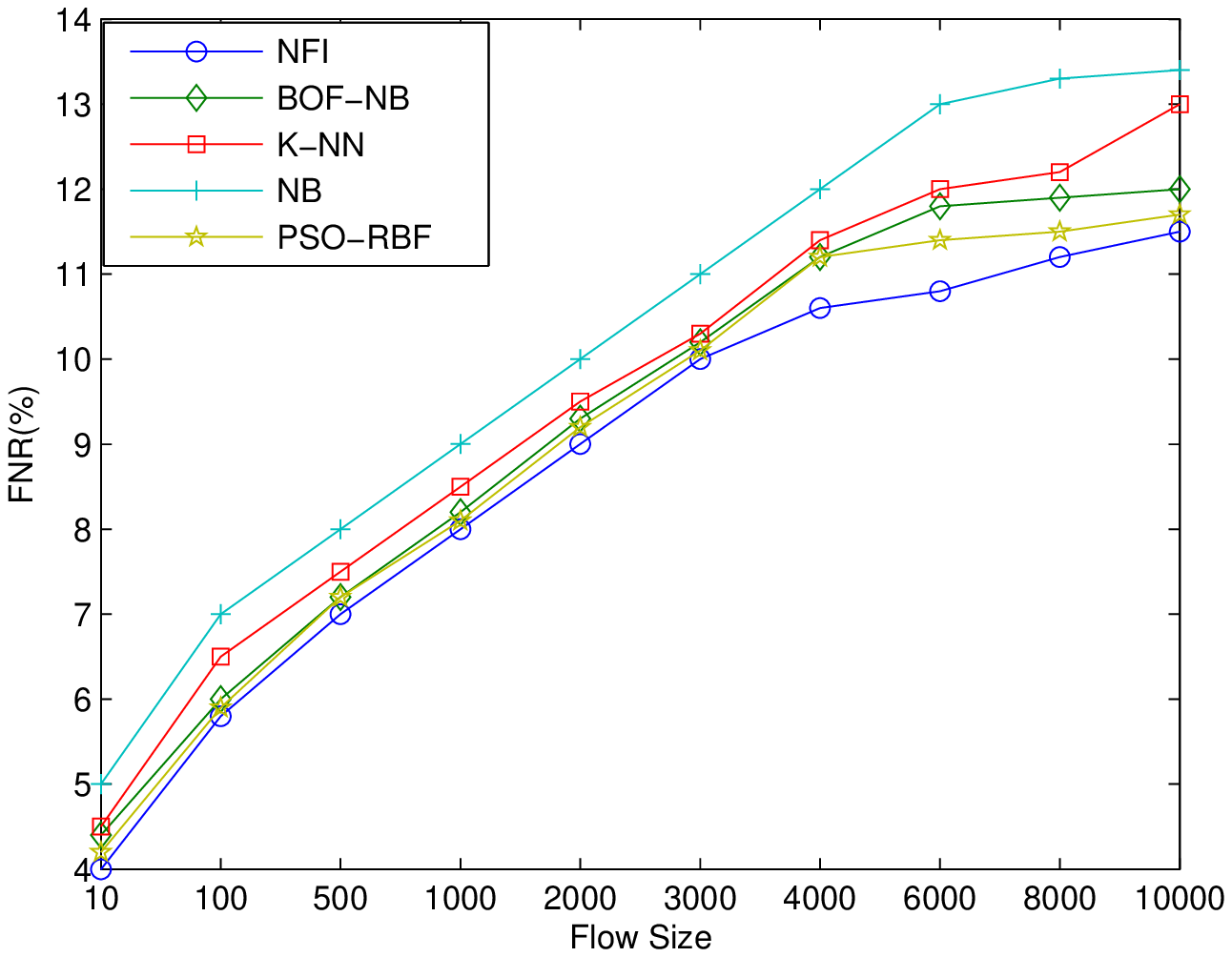}}
\centerline{\footnotesize\begin{tabular}{c} Fig.\ 7.\ Performance evaluation in different flow size
\end{tabular}}
\label{fig5}
\end{figure*}
\begin{figure*}[htb]
\centering
\subfigure[FPR in different flow duration] {\includegraphics[height=2.18in,width=3.18in]{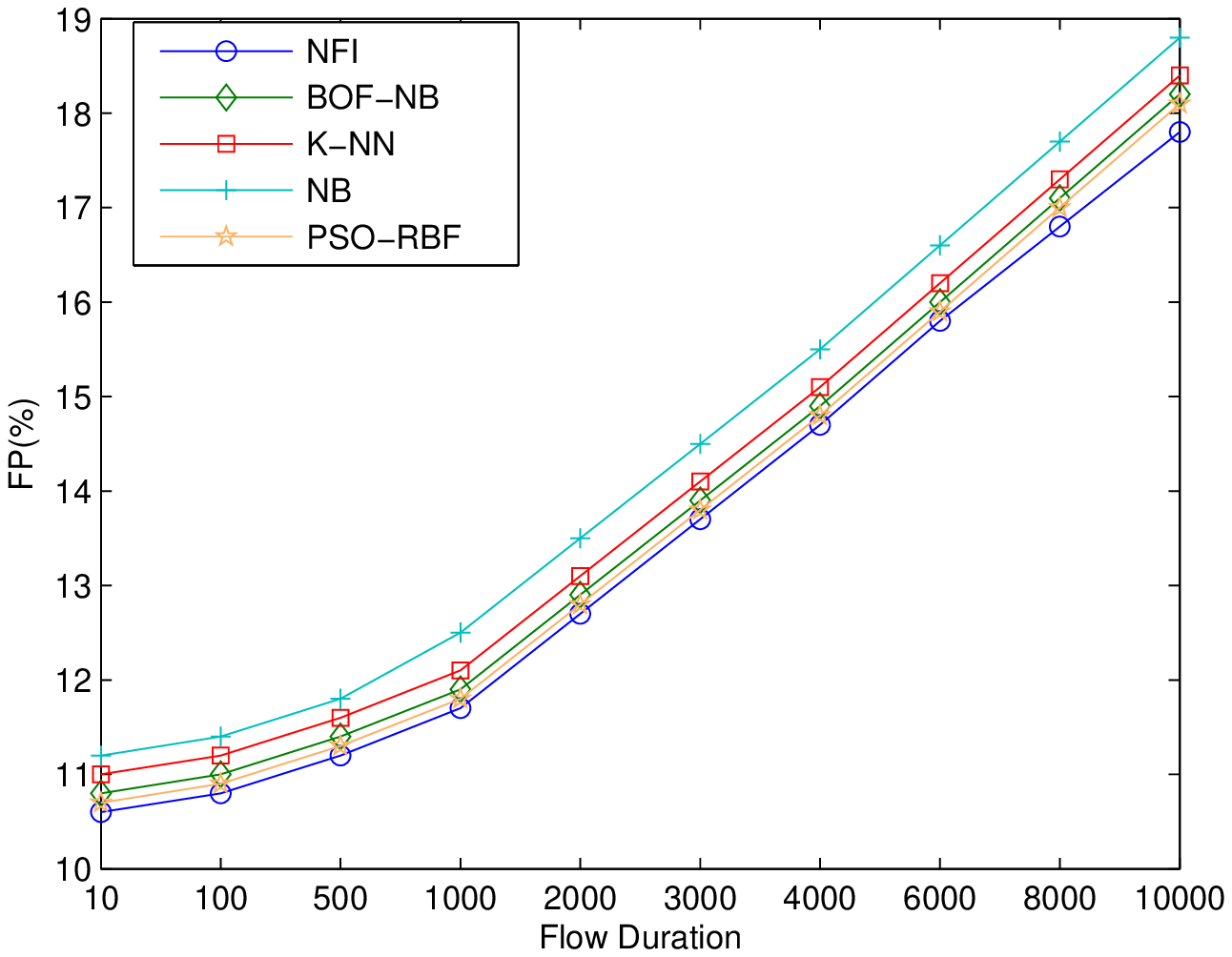}}
\subfigure[FNR in different flow duration] {\includegraphics[height=2.18in,width=3.18in]{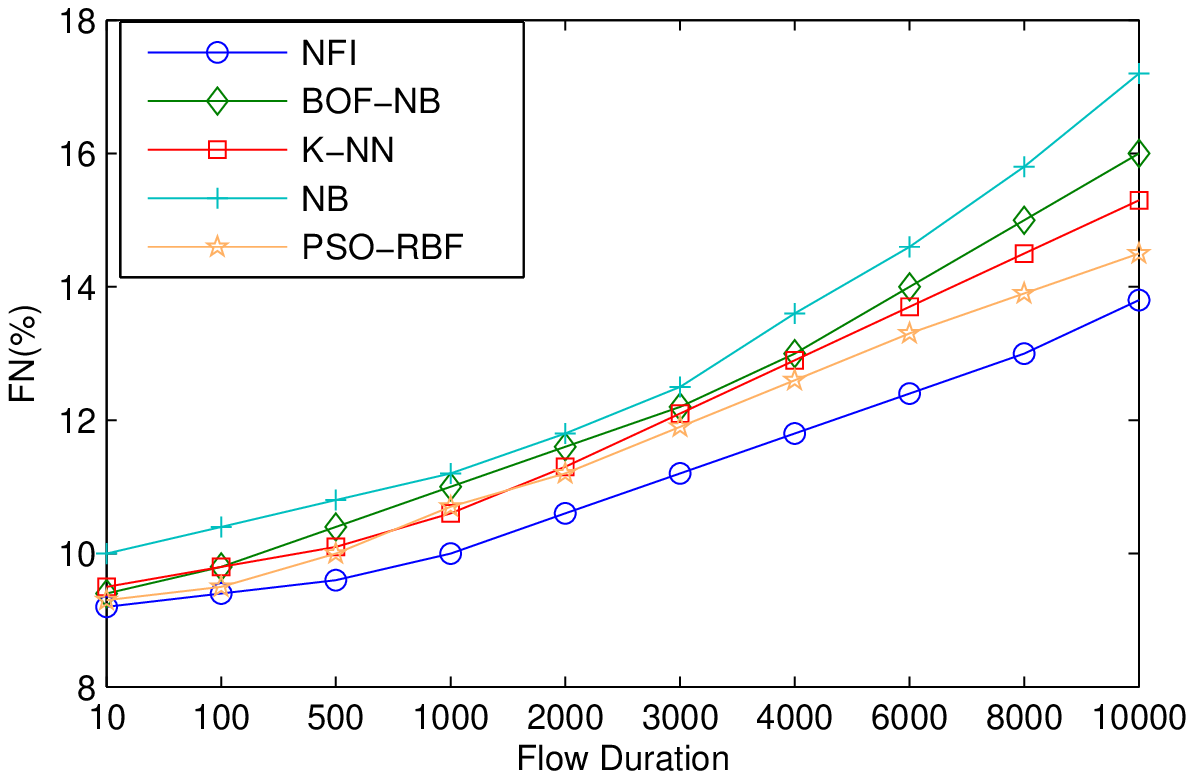}}
\subfigure[TPR in different flow duration] {\includegraphics[height=2.18in,width=3.18in]{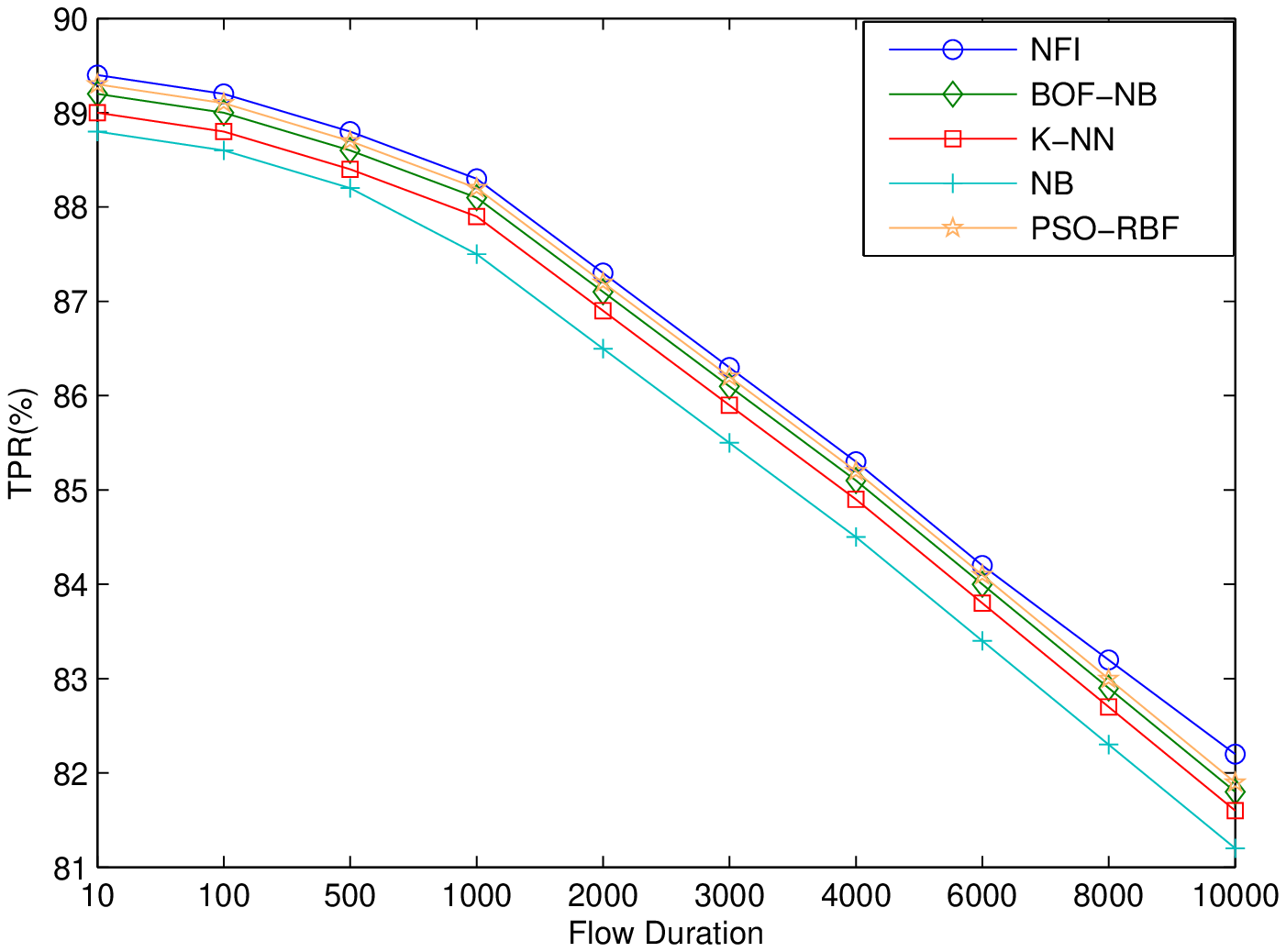}}
\subfigure[TNR in different flow duration] {\includegraphics[height=2.18in,width=3.18in]{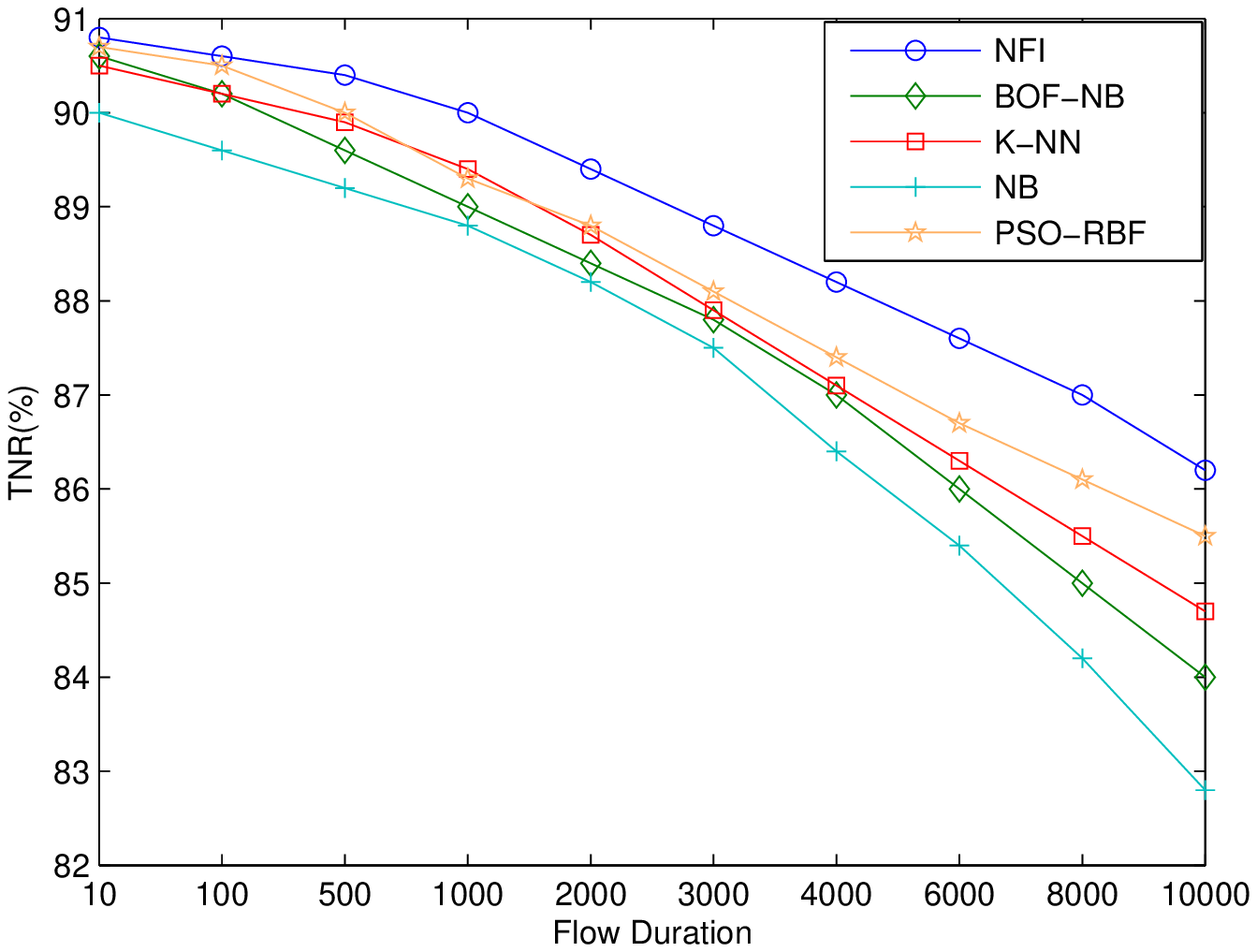}}
\centerline{\footnotesize\begin{tabular}{c} Fig.\ 8.\ Performance evaluation in different flow duration
\end{tabular}}
\label{fig5}
\end{figure*}
The experimental results indicate that different sampling ratios exerted different impacts on features. When the sampling ratio was 1:256, the flow features clearly changed, although flow behavior was similar to that with no sampling. Thus, 1:256 was chosen as the default sampling strategy in the following experiment.

\subsection{Experimental results and analysis}
This paper considered the full 16 features from the SKYPESET, and the Symmetrical UncertAttributeSetEval (SU) and FCBF evaluation methods were adopted. The identification method used NBI and NBK, and a 10-fold cross-validation method was used for algorithm evaluation. The 10-fold cross-validation method is commonly used for precision with the premise that a dataset is divided into 10 parts, nine of which constitute the training data with one representing the test data. Each experiment obtained accuracy measurements, the average of 10 accuracy iterations for the algorithm was considered the total accuracy. The feature selection algorithm was used to reduce features; cross-validation was adopted to evaluate the classifier. Experiment results appear in Fig. 6-8 and Table 4.

Table 4 shows that the FCBF feature selection algorithm was used to sort FCBF values and compare the smallest FCBF value in the feature column and the maximum SU(Ai,Aj) in the feature column; if the values were identical, they were deleted. Only the larger feature column of the FCBF was preserved, and then the FCBF algorithm optimized and combined the features column. Theoretically, these features should exert a larger impact on the overall accuracy of flow F. The overall accuracy of the FCBF algorithm demonstrated the highest accuracy rate because it was used to select three features (numbers 9, 12, and 14) from 16 features in the SKYPE\-SET, and the classifier was more independent; detailed information is presented in Table 3.
\begin{table}
{\tabcolsep=2.5pt \footnotesize
\label{tab:1}       
\begin{center}
\begin{tabular}{cccc}
\multicolumn{4}{c}{\bf Table 4.\ Optimal Feature}\\
\hline\noalign{\smallskip}
\centering
ID& feature abbreviation & feature describe & SU \\
\noalign{\smallskip}\hline\noalign{\smallskip}
9& Lport & low port & 0.7112518\\
12& BPS & bytes/duration & 0.2498672\\
14& Doublepktratio & Biodirection packets length ratio & 0.1655624\\
\noalign{\smallskip}\hline
\end{tabular}
\end{center}
}
\end{table}

This paper proposed the NFI method based on flow records compared to three state-of-the-art methods: the naive Bayes method, K-NN, Jun Zhang’s BOF-NB method \cite{Zi2014Boosted} and Yong Chen's PSO-RBF method\cite{Zi2015Boosted}. Traffic identification based on flow records with few features had nearly the same identification results as full-packet data, providing a sound means of online traffic classification. By adding a few of the above-mentioned metric features to NETFLOW and building new metric features, better classification results could be achieved along with improved online classification and identification.

%

Figs. 6-8 show that FPR, FNR, TPR, and TNR are different with regard to flow length, size, and duration. When the training sample increased, so did the FPR and FNR. The proposed NFI method was less than the  naive Bayes, KNN, PSO-RBF, and BOF-NB methods. Though the BOF-NB approach had a similar FNR to NFI in the 10–3000 range, the difference in FNR became more obvious with an increasing flow size. TPR and TNR decreased, and the proposed method was greater than the na?ve Bayes, K-NN, PSO-RBF, and BOF-NB approaches; the NFI method adopts an updating mechanism to update the new data sample, thus improving identification accuracy and reducing false positives and false negatives.

\begin{table}[h]
\center
\begin{tabular}{llllll}
\multicolumn{4}{c}{\bf Table 5.\ Performance evaluation}\\
\hline\noalign{\smallskip}
Algorithm & Precision & Recall & OA & F-measure \\
\noalign{\smallskip}\hline\noalign{\smallskip}
NFI & 93.6\% & 94.0\% & 96.7\% & 94.6\% \\
NaiveBayes & 91.3\% & 92.5\% & 93.2\% & 91.9\% \\
K-NN & 92.2\% & 92.8\% & 93.8\% & 92.5\% \\
PSO-RBF & 93.2\% & 93.8\% & 95.9\% & 93.5\% \\
BOF-NB & 92.8\% & 93.4\% & 94.8\% & 93.2\% \\
\noalign{\smallskip}\hline
\end{tabular}
\end{table}
Table 5 presents the final results obtained after analyzing 760 packets. Results were quite good for most Skype flows. File uploads and downloads were easily distinguished based on the flow attribute combining the direction with the packet size distribution (cf. attribute M8 in Table 1). Classification was based on the fact that the packet sizes sent from the client differed significantly from those sent in the opposite direction.
\section{Conclusions}
Traffic identification is a core issue in network traffic planning and management. This paper obtained network data and adopted an L7-filter to label data and construct a baseline SKYPE\_SET data set. Results show that the feature selection algorithm proposed in this paper can achieve better classification results and a higher identification rate than other methods. Contributions of this paper are as follows: (1) construction of a standard SKYPE-SET dataset; (2) proposed feature metrics based on flow; (3) NFI based on an updating mechanism was applied in Skype identification. Based on this study, our subsequent work will involve further evaluation of flow data and flow measure features to provide data support for future research. Upcoming work will also seek to improve the feature selection algorithm to select better metrics and propose better metric features.\cite{rajahalme2011ipv6} and \cite{levandoski2008application} and \cite{hall2009weka}.
\section{Acknowledgments}
This paper is supported by Project supported by the National Natural Science Foundation of China (Grant No.U1504602), China Postdoctoral Science Foundation(Grant No.2015M572141), National 973 Plan Projects (Grant No.2009CB320505) and National Science and Technology Plan Projects (Grant No.2008BAH37B04), and Education Department of Henan Province Science and Technology Key Project Funding(Grant No.14A520065). The authors would like to thank Universidad Autonomade Madrid, Northeast China Center of CERNET for providing their datasets.
\section{Conflict of Interest Statement}
The authors of the paper immediately below certify that they have no affiliations with or involvement in any organization or entity with any financial interest.

\bibliographystyle{unsrt}
\bibliography{references}


\end{document}